\documentclass[preprint,12pt,number]{elsarticle}
\usepackage[a4paper,top=2.5cm,bottom=3.2cm,left=2.5cm,right=2.5cm]{geometry}
\usepackage{framed} 
\usepackage{nomencl} 
\usepackage{multirow}
\usepackage{multicol}
\usepackage{tabularx}

\biboptions{sort&compress}
\usepackage{amssymb}
\usepackage{amsmath}
\usepackage{graphicx}
\usepackage{graphics}
\usepackage{subcaption}
\usepackage{nomencl}
\usepackage{float}
\usepackage{epstopdf}
\usepackage{epsfig}
\usepackage{booktabs}
\usepackage{caption}
\usepackage{caption}
\usepackage{array}
\usepackage{float}    
\usepackage{xcolor}
\usepackage{hyperref}

\hypersetup{
    colorlinks=true,
    linkcolor=blue,
    citecolor=red,
    urlcolor=magenta
}

\makenomenclature

\biboptions{sort&compress}
\usepackage{amssymb}
\usepackage{amsmath}
\usepackage{graphicx}
\usepackage{graphics}
\usepackage{subcaption}
\usepackage{nomencl}
\usepackage{float}
\usepackage{epstopdf}
\usepackage{epsfig}
\usepackage{booktabs}
\usepackage{caption}
\usepackage{multirow}
\usepackage{caption}
\usepackage{array}
\usepackage{float}    
\usepackage{xcolor}
\usepackage{hyperref}

\hypersetup{
    colorlinks=true,
    linkcolor=blue,
    citecolor=red,
    urlcolor=magenta
}

\DeclareUnicodeCharacter{2082}{\ensuremath{_2}}

\makeatletter
\newcounter{savesection}
\newcounter{apdxsection}
\renewcommand\appendix{\par
  \setcounter{savesection}{\value{section}}%
  \setcounter{section}{\value{apdxsection}}%
  \setcounter{subsection}{0}%
  \gdef\thesection{\@Alph\c@section}}
\newcommand\unappendix{\par
  \setcounter{apdxsection}{\value{section}}%
  \setcounter{section}{\value{savesection}}%
  \setcounter{subsection}{0}%
  \gdef\thesection{\@arabic\c@section}}
\makeatother

\begin{document}
\begin{frontmatter}
\title{Aggregation Effects on Heat Transfer in Viscoplastic Nanofluid Entrance Flows}
\author[inst1]{Deepa Madivalar\corref{cor1}}
\ead{d.227ma009@nitk.edu.in}
\cortext[cor1]{Corresponding author}

\author[inst1]{Vishwanath Kadaba Puttanna}
\author[inst1]{A Kandasamy} 

\affiliation[inst1]{organization={Department of Mathematical and Computational Sciences},
            addressline={National Institute of Technology Karnataka, Surathkal}, 
            city={Mangalore},
            postcode={ 575025}, 
            state={Karnataka},
            country={India}}

\begin{abstract}
This study numerically investigates heat transfer enhancement in laminar, incompressible viscoplastic nanofluid flow through the entrance region of a circular cylinder with a uniformly heated wall, including the effects of both, non-aggregation and aggregation of nanoparticles. Nanofluid properties are modeled using Brinkman and Maxwell models in the case of non-aggregation, and Krieger–Dougherty, Maxwell–Bruggeman models in the case of aggregation, while the viscoplastic behavior is described by the Bingham–Papanastasiou model. The governing boundary layer equations are solved using a finite-difference method. The effects of yield stress and nanoparticle volume fraction (up to 5\%) on friction, pressure drop, and Nusselt number are analyzed, and performance evaluation criteria are evaluated to identify the optimal volume fraction for maximum efficiency.

\end{abstract}

\end{frontmatter}

\section{Introduction}
The enhancement of heat transfer has become a key focus in modern research, with nanofluids emerging as a promising solution. Nanofluids are base fluids that contain a stable suspension of nanoparticles \cite{choi1995enhancing}, which can markedly enhance thermal transport. These fluids find extensive applications in heat exchangers, petrochemical industries, electronic cooling systems, and biopharmaceutical processes ~\cite{saidur2011review, sadeghinezhad2016comprehensive}. Many studies have demonstrated that the superior thermal performance of nanofluids arises from the much higher thermal conductivity of the dispersed nanoparticles compared to that of the base fluid ~\cite{dai2024mechanism,ma2023heat}.

The flow of nanofluids within a pipe/annuli plays a vital role in optimising the design of heat exchangers and automotive cooling systems, thereby enhancing thermal performance and overall efficiency \cite{ali2024effect}.
Numerous experimental and numerical studies have investigated nanofluids formulated with both Newtonian \cite{heris2006experimental,maiga2005heat,teamah2018influence,najafabadi2024entry,muhammad2025flow} and non-Newtonian base fluids \cite{akbari2017effect, hazeri2021three,venthan2019analysis,venthan2021theoretical,ouyahia2017numerical,ahmed2025numerical,hussain2025thermal}, combined with various nanoparticles such as $Cu$, $Ag$, $Au$, $SiO_2$, $Al_2O_3$, $TiO_2$, and $CuO$. Among these nanoparticles, the most commonly used nanoparticles are $Al_2O_3$ and $Cu$.

Many researchers have analyzed the flow in a pipe numerically by incorporating appropriate single-phase nanofluid models based on their thermophysical properties. In 2018, Teamah et al. \cite{teamah2018influence} examined the developing flow region of a pipe using water-based nanofluids containing up to 10 \% volume fraction of three different nanoparticles: $Al_2O_3$, $TiO_2$, and $Cu$. A numerical single-phase analysis with the Brinkman viscosity model was performed assuming constant thermophysical properties of the nanofluids. The results indicated that adding nanoparticles did not affect the viscous boundary layer but did alter the thermal boundary layer due to changes in viscosity and thermal conductivity; however, both viscous and thermal boundary layers were influenced by variations in Reynolds number. Their comparative study further showed that  $Cu$ nanoparticles provided the greatest enhancement in heat transfer among the three. Later, in 2024, Najafabadi et al. \cite{najafabadi2024entry} numerically studied  $Al_2O_3$-water nanofluid with up to 5\% volume fraction in the pipe entrance region with uniform heat flux, applying the Klazly–Bognar viscosity model and the Maxwell thermal conductivity model, where the thermophysical properties were assumed to be temperature dependent. Their results showed that, for a given Reynolds number, the velocity entry length decreases with increasing nanoparticle volume fraction due to the rise in viscosity, whereas the thermal entry length increases as the nanoparticle volume fraction increases. In 2025, Muhammad et al. \cite{muhammad2025flow} studied the flow of hybrid nanofluid in a coaxial cylinder using the Kellar-box method. They considered the effects of the magnetic field, slip velocity, and thermal slip conditions at the inner cylinder. Their results revealed that there is an increase in the skin friction factor due to enhanced momentum transfer.

 It is observed that all the aforementioned studies assumed nanofluids to behave as Newtonian fluids, modeling them as homogeneous mixtures under the single-phase approach. However, in many cases, either the addition of nanoparticles to the base fluid alters its rheological characteristics, causing it to exhibit non-Newtonian behavior \cite{labib2013numerical} or the base fluid itself can be a non-Newtonian fluid.
In 2013, Labib et al. \cite{labib2013numerical} demonstrated this by using a hybrid nanofluid consisting of carbon nanotubes (CNTs) and $Al_2O_3$ in water numerically using a two-phase model. Their results showed that the hybrid nanofluid was more effective, primarily due to the non-Newtonian shear-thinning behaviour of the CNT-based nanofluid, which had a significant impact, especially in the entrance region of the flow domain.

However, the entrance-region flow of non-Newtonian fluids, especially viscoplastic fluids, plays a crucial role due to the presence of yield stress. Viscoplastic fluids have received significant attention in the heat transfer analysis, which is affected by the yield stress of the fluid \cite{nadiminti2020heat}. Venthan et al. \cite{venthan2019analysis} investigated viscoplastic nanofluid flow in the entrance region of a concentric annulus using an ideal Bingham model with silver ($Ag$) and copper ($Cu$) nanoparticles, with the assumption
that the stress is above the yield stress everywhere in the flow region.
Their study focused on velocity and pressure drop in this region. They found that at lower nanoparticle volume fractions, silver and copper nanofluids exhibit similar velocity trends; however, at higher volume fractions, the silver nanofluid shows greater velocity, while the copper nanofluid experiences a higher pressure drop compared to the silver nanofluid.
Later, Venthan et al. \cite{venthan2021theoretical} numerically investigated heat transfer enhancement of Bingham fluids in an annulus using an ideal Bingham model with $Al_2O_3$ and $TiO_2$ nanoparticles. Their results showed that the 
$TiO_2$-Bingham nanofluid provides greater heat transfer enhancement than the 
$Al_2O_3$-Bingham nanofluid, and that a larger annular gap accelerates heat transfer.

Most of the available literature mainly addresses the fully developed region of the flow domain. However, in real systems such as pipes and annuli, the entrance region is equally important due to the progressive development of velocity and thermal boundary layers. In this developing region, nanofluids show pronounced changes in flow behaviour, where nanoparticle concentration significantly affects key parameters like velocity, pressure, and temperature. 
Although some studies have examined this region for Newtonian nanofluids, few studies have examined using viscoplastic nanofluids.
The present study focuses on the flow of a viscoplastic nanofluid in the entrance region of a circular straight cylinder/pipe. 

In a nanofluid flow, inter-particle interactions and random nanoparticle motion cause nanoparticles in a nanofluid system to aggregate, which significantly deviates from the behaviour of uniformly dispersed (i.e., non-aggregated) nanoparticles, especially when the volume fractions are high. The effect of aggregation significantly alters the thermophysical properties of nanofluids by increasing the effective viscosity and thermal conductivity and thereby enhancing heat transfer through conductive pathways. Some of the studies considered this effect of aggregation on non-Newtonian fluids for different geometries, such as curved surface, rotating disk, stretchable cylinder \cite{srilatha2023heat,alsulami2023three,sarma2025comparative,jan2025enhanced}. In their studies(see in \autoref{Table:Lit_KD}), they analyzed the impact of non-aggregated and aggregated nanoparticle effects on the flow behaviour using a single-phase homogeneous approach with the Krieger–Dougherty viscosity model combined with the Maxwell–Bruggeman thermal conductivity model and their studies reveal that aggregation intensifies heat transfer characteristics while increasing flow resistance. In the entrance region, most studies focus on uniformly dispersed nanoparticles, using the Brinkman model for viscosity and the Maxwell model for thermal conductivity. However, there is a lack of research that takes into account the effects of nanoparticle aggregation in this region.

\begin{table}[htbp]
\caption{Some previous works utilized the Krieger-Dougherty and Maxwell-Bruggeman models for aggregated nanoparticles in non-Newtonian fluids across different geometries.}
\label{Table:Lit_KD}
\centering
\resizebox{\textwidth}{!}{
\begin{tabular}{|c|c|c|c|}
\hline
\textbf{Author (s)} & \textbf{Geometry}  & \textbf{Basefluid}  &\textbf{Classification of flow }  \\ [4pt]
\hline
Srilatha et.al \cite{srilatha2023heat} 
& Rotating disk 
& Maxwell fluid 
& Rotating flow\\[4pt]
\hline


Shen et.al \cite{shen2024entropy} 
& Thin film 
& Micropolar
& Electrically conducting flow \\[4pt]
\hline

Sarma et.al \cite{sarma2025comparative} 
& Inclined stretching surface 
&  Boger fluid
& MHD flow  \\[4pt]
\hline

Jan et.al
\cite{jan2025enhanced} 
& Curved surface 
& Viscoplastic 
& Porous medium \\[4pt]
\hline

Anitha et.al \cite{anitha2025impact} 
& Inclined stretching surface 
&  Boger hybrid
& MHD flow \\[4pt]
\hline
\textbf{Present Work}&\textbf{Circular cylinder}&\textbf{Viscoplastic}&\textbf{Entrance region}\\
\hline
\end{tabular}
}
\end{table}

\subsection{Novelty}
A comprehensive review of the literature reveals that the heat transfer and flow in the entrance region of viscoplastic nanofluids in circular cylinders, comparing the effects of non-aggregation and aggregation of nanoparticles employing the single-phase models, have not been investigated to the best of our knowledge (see \autoref{Table:Literature}). In the present work, we have numerically investigated the flow and heat transfer characteristics of a viscoplastic nanofluid in the entrance-region of a uniformly heated circular cylinder using a single-phase Brinkman model for viscosity and Maxwell model for conductivity in the case of non-aggregation and Krieger–Dougherty model for viscosity and Maxwell-Bruggeman model for conductivity in the case of aggregation \cite{shen2024entropy}. The viscoplastic flow behaviour is described using the Bingham-Papanastasiou model, and the flow equations are governed by the Prandtl boundary-layer assumptions. The analysis accommodates nanoparticle volume fractions up to $5\%$.

\begin{table}[htbp]
\caption{Comparison of the present study with previous research on different nanofluid models in pipe/annuli geometry with single phase/two phase models.}
\label{Table:Literature}
\centering
\resizebox{\textwidth}{!}{
\begin{tabular}{|c|c|c|c|c|c|}
\hline
\textbf{Author (s)} & \textbf{Base Fluid}  & \textbf{Viscosity} & \textbf{Assumption} & \textbf{Entrance}& \textbf{Aggregation}  \\ [2pt]
 &  &\textbf{model} &  &  \textbf{region}&   \\ [4pt]
\hline
Teamah et al. \cite{teamah2018influence}  & Newtonian  & Brinkman  & Constant wall temp  & Yes  & No  \\[4pt]
\hline
 Najafabadi et al. \cite{najafabadi2024entry}  & Newtonian  & Klazly-Bognar  & Constant heat flux & Yes & No \\[4pt]
 \hline
Venthan et al. \cite{venthan2021theoretical} & Viscoplastic & Brinkman & Adiabatic and isothermal & Yes & No \\[4pt]
\hline
Hazeri et al. \cite{hazeri2021three} & Power-law & Wang model & Constant heat flux & No & No \\[4pt]
\hline
\textbf{Present Study} & \textbf{Viscoplastic} & \textbf{Kriger-dougherty} & \textbf{Constant wall temp} & \textbf{Yes} & \textbf{Yes} \\
 &  & \textbf{\& Brinkman }&  &  &  \\
\hline
\end{tabular}
}
\end{table} 
\subsection{Applications} 
The flow and heat transfer characteristics of viscoplastic nanofluids in the entrance region of a cylindrical system are important in many engineering applications, such as heat exchangers, microfluidic devices, cooling towers, biomedical applications like targeted drug delivery, polymer processing, paint, and high-performance machinery. It also plays a significant role in controlling the rheology of drilling mud during the process of drilling petroleum wells, which exhibit highly non-Newtonian fluid behaviour. When designing drilling machines for petroleum extraction, it is essential to understand the flow behaviour at the entrance and optimise the properties accordingly. Additionally, maintaining pressure is crucial to prevent issues while drilling for petroleum \cite{soares1999heat}.
The combined influence of yield stress and nanoparticle concentration leads to enhanced thermal conductivity, particularly in the inlet region where the boundary layers are developing. These features make the flow in the entrance region highly advantageous for improving efficiency in industrial transport processes involving viscoplastic fluids when the flow is through a short pipe, where the velocity and temperature remain in a developing state.

The present study includes analysis of hydrodynamic and thermal entry regions, considering the effects of yield stress (i.e., Bingham number) and nanoparticle volume fractions (up to $5\%$) on velocity, friction coefficient, temperature and Nusselt number. The enhancement of the heat transfer of viscoplastic nanofluids in the case of non-aggregated and aggregated conditions is analyzed. The overall efficiency of the non-aggregation and aggregation-based nanofluid models is discussed through the calculations of Performance Evaluation Criteria.

\section*{Nomenclature}
\begin{center}
\fbox{
\begin{minipage}{0.95\textwidth}
\small
\begin{multicols}{2}

\textbf{Variables}\\
$r,x$ : Cylindrical coordinates (m) \\
$r_1$ : Radius of cylinder (m)\\
$\tilde{v}_x, \tilde{v}_r$ : Axial and radial velocity (m/s) \\
$\tilde{T}$ : Temperature (K) \\
$T_b$ :  Bulk temperature(K)\\
$T_w$ : Wall temperature(K)\\
$q_w$ : Heat flux (W/m$^2$)\\
$p$ : Pressure  (kg ms$^{-2}$)\\
$p_0$ : Initial pressure (kg ms$^{-2}$)\\
$u_0$ : Initial axial velocity(m/s)\\
$T_0$ : Initial temperature(K)\\
$\tau_{rx}$: Shear stress in the x-direction\\ and perpendicular to the r-direction (Pa) \\
$\tau_0 $: Yield stress(Pa)\\
$\dot\gamma$: Shear rate \\
$m$: Regularization parameter (s)\\\\
\textbf{Dimensionless Variables}\\
$\eta,\xi$: Dimensionless cylindrical coordinates\\
$U,V $: Dimensionless axial, radial velocities\\
$\Theta$: Dimensionless temperature\\
$\Theta_b$: Dimensionless bulk temperature\\

\textbf{Dimensionless parameters}\\
$M$: Dimensionless regularization parameter\\
$B_n$: Bingham number\\
$Nu$: Nusselt number\\
$R_e$: Reynolds number\\
$Pr$ : Prandtl number\\

\textbf{Nanofluid}\\
$k$ : Thermal conductivity (W/mK) \\
$\alpha$ : Thermal diffusivity(m$^{2}$/s)\\
$\rho$ : Density (kg/m$^3$) \\
$\mu$ : Apparent viscosity (Pa·s) \\
$\mu_b$ : Plastic viscosity (Pa·s) \\
$c_p$ : Specific heat capacity(J\,kg$^{-1}$K$^{-1}$) \\
$\phi$ : Nanoparticle volume fraction \\
$R$ : Particle radius \\
$D$ : Fractal index \\

\textbf{Subscripts}

$nf$: Nanofluid \\
$bf$: Base fluid \\
$agg$: Aggregate \\
$p$ : Particle\\
\end{multicols}
\end{minipage}
}
\end{center}

\section{Mathematical Model}
In this section, we describe the problem of interest along with governing equations, including the appropriate constitutive equations and associated boundary conditions.
\subsection{\textbf{Problem statement}}
\begin{figure}[H]
    \centering
    \includegraphics[width=0.9\linewidth]{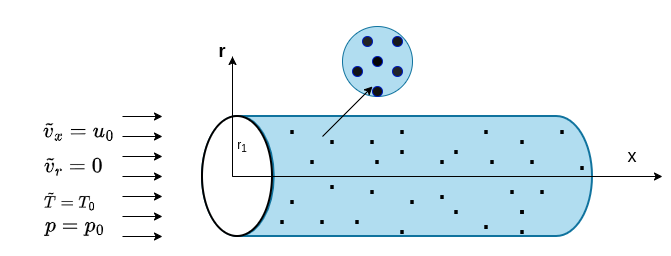}
    \caption{Schematic of the viscoplastic nanofluid flow in a cylinder/pipe}
    \label{fig:Cylinder}
\end{figure}

We consider a viscoplastic nanofluid flow through a circular straight cylinder of radius $r_1$ with a uniform wall temperature as shown in \autoref{fig:Cylinder}. The flow is assumed to be steady, incompressible, and laminar. A cylindrical coordinate system with $(r,\theta,x)$ is placed in the centre of the inlet such that $r$ measures the distance in the radial direction, $\theta$ is the azimuthal angle measuring the rotation of the radius vector, and $x$ denotes the axial length of the cylinder. At the inlet, the fluid enters with a uniform velocity $u_0$, constant pressure $p_0$ maintained at a constant temperature $T_0$.  The working fluid is a viscoplastic nanofluid, which is a homogeneous mixture of viscoplastic fluid as the base fluid and suspended nanoparticles. The suspended nanoparticles are studied under both non-aggregated and aggregated conditions. In the next subsection,  (\ref{Section:Thermo_physical_Prop}) we describe the details of the thermophysical properties adopted in our analysis.

\subsection {\textbf{Thermophysical properties of nanofluid}}
\label{Section:Thermo_physical_Prop}
A typical example of nanoparticles and base fluid are $Al_2O_3$ and Carbopol solution respectively and their thermophysical properties are as shown in \autoref{tab:Thermo_Prop}. Thermophysical properties of the nanofluid are considered as constant which are given in the \autoref{tab:Thermo_models_agglomeration}
The effective density, viscosity, specific heat,  and thermal conductivity of the nanofluid are represented by \(\rho_{nf}\), \(\mu_{nf}\), $(c_p)_{nf}$, and \(k_{nf}\), respectively. The effective viscosity of the nanofluid is calculated using the the Brinkman model for the non-aggregation case and Krieger–Dougherty model for the aggregation case. In contrast, the effective thermal conductivity is evaluated based on Maxwell model for non-aggregation case the Maxwell–Bruggeman model for aggregation case, as summarized in \autoref{tab:Thermo_models_agglomeration} for both aggregation and non-aggregation scenarios.
\begin{table}[htbp]
    \renewcommand{\arraystretch}{1.2}
\centering
\caption{Properties of nanoparticles and basefluid ~\cite{teamah2018influence}}
\label{tab:Thermo_Prop}
\begin{tabular}{|l| l |l |l |}
\hline
Material & Density & Specific heat  &  Thermal conductivity\\
\hline
$Al_2O_3$ & 3970 & 765 &40\\


Base fluid & 997.1 & 4179  & 0.613\\
\hline
\end{tabular}%
\end{table}

\begin{table}[htbp]
\centering
\caption{Effective models for determining thermophysical properties of nanofluids with and without aggregation~\cite{shen2024entropy,bhavya2025computational}}
\label{tab:Thermo_models_agglomeration}

\resizebox{\textwidth}{!}{%
\renewcommand{\arraystretch}{1.8}
\begin{tabular}{|l|l|l|}
\hline
\textbf{Property} & \textbf{Non-Aggregation} & \textbf{Aggregation} \\
\hline
Density &
$\rho_{nf} = (1 - \phi)\rho_{bf} + \phi\rho_p$ &
$\rho_{nf} = (1 - \phi_{agg})\rho_{bf} + \phi_{agg}\rho_{agg}$ \\[4pt]
Viscosity &
$\mu_{nf} = \mu_{bf} (1-\phi)^{-2.5}$&
$\mu_{nf} = \mu_{bf} \left(1-\dfrac{\phi_{agg}}{\phi_{max}}\right)^{-2.5\phi_{max}}$ 
 \\[4pt]
Specific heat &
$(\rho c_p)_{nf} = (1 - \phi)(\rho c_p)_{bf} + \phi(\rho c_p)_p$ &
$(\rho c_p)_{nf} = (1 - \phi_{agg})(\rho c_p)_{bf} + \phi_{agg}(\rho c_p)_{agg}$ \\[8pt]
Thermal conductivity &
$k_{nf} = k_{bf}\dfrac{k_p + 2k_{bf} - 2\phi (k_{bf} - k_p)}%
{k_p + 2k_{bf} + \phi (k_{bf} - k_p)}$ &
$k_{nf} = k_{bf} \dfrac{k_{agg} + 2k_{bf} - 2\phi_{agg}(k_{bf} - k_{agg})}%
{k_{agg} + 2k_{bf} + \phi_{agg} (k_{bf} - k_{agg})}$ \\[8pt]

\hline
\end{tabular}%
}
\end{table}

Further \(\rho_{bf}\) and  \(\rho_p\) denote the density of the base fluid and the nanoparticles, respectively; \((c_p)_{bf}\) and \((c_p)_{p }\) 
 represent their specific heat capacities; and \(k_{bf}\) and \(k_{p} \) 
 correspond to the thermal conductivity of the base fluid and the nanoparticles, respectively. The viscosity of the base fluid \(\mu_{bf}\), according to the Bingham-Papanastasiou model \cite{article}, is expressed as follows:
 
\begin{eqnarray}
\mu_{bf}&=&\left[\mu_{b}+\frac{{\tau}_{0}[1-exp(-m|\frac{\partial \tilde{v}_x}{\partial r}|)]}{|\frac{\partial \tilde{v}_x}{\partial r}|}\right]
\end{eqnarray}
where $\mu_b$ is the plastic viscosity and $\tau_0$ is the yield stress of the fluid, and m is the regularization parameter. 

The effective thermophysical properties of the aggregated nanoparticles are evaluated using the following relations:
\begin{align}
\rho_{agg} &= \rho_{bf}(1 - \phi_{in}) + \rho_p \phi_{in}, \label{eq:rho_agg} \\
(\rho c_p)_{agg} &= (1 - \phi_{in})(\rho c_p)_{bf} + \phi_{in}(\rho c_p)_p, \label{eq:cp_nf}
\end{align}

The effective thermal conductivity of the aggregate nanofluid is determined as~\cite{shen2024entropy}:
\begin{align}
k_{agg} &= \frac{k_{bf}}{4} \Biggl[
\left\{ (3\phi_{in} - 1)\frac{k_p}{k_{bf}} + 3(1 - \phi_{in}) - 1 \right\} \nonumber \\
&\quad + \left(
\left\{ (3\phi_{in} - 1)\frac{k_p}{k_{bf}} + 3(1 - \phi_{in}) - 1 \right\}^2 
+ 8\frac{k_p}{k_{bf}}
\right)^{1/2}
\Biggr]. \label{eq:k_agg}
\end{align}

The effective nanoparticle volume fraction within aggregates is given by \cite{bhavya2025computational}:
\begin{align}
\phi_{agg} &= \frac{\phi}{\phi_{in}},~~\text{and}\quad \phi_{in} = \left( \frac{R_{agg}}{R_p} \right)^{D - 3}  \\
D=&1.8,\quad \left( \frac{R_{agg}}{R_p} \right)=3.34,\quad \phi_{max}=0.605\label{eq:phi_in} \nonumber
\end{align}

where $\phi$ and  $\phi_{agg}$  denote the volume fraction of non-aggregated and aggregated nanoparticles, and $\phi_{in}$ represents the nanoparticle volume fraction within an aggregate. 
Here, $D$ is the fractal index, $R_{agg}$ is the aggregate radius, and $R_p$ is the radius of a primary nanoparticle.

\subsection{\textbf{ Problem formulation}}
The flow is modeled using the governing equations under the assumptions of Prandtl's boundary layer theory, considering incompressible, steady and laminar conditions. For a viscoplastic-based nanofluid, the base fluid and nanoparticles are in thermal equilibrium, viscous dissipation is neglected and the continuity, x-momentum, and energy equations are given as follows \cite{baioumy2021bingham,venthan2021theoretical},

\noindent \textbf{Continuity equation:}
\begin{equation}
    \frac{\partial \tilde{v}_x}{ \partial x}+\frac{1}{r}\frac{\partial (r\tilde{v}_r)}{\partial r}=0 \label{eqn:Continuity_eqn}
\end{equation}

\noindent \textbf{Momentum equation:}

\begin{equation}
   \rho_{nf}\left[\tilde{v}_x\frac{\partial \tilde{v}_x}{\partial x}+ \tilde{v}_r\frac{\partial \tilde{v}_x}{\partial r}\right]=-\frac{dp}{dx}+\frac{1}{r}\frac{\partial }{\partial r}\left({r\tau_{rx}}\right)\label{Eq:mom} 
\end{equation}

Where $\tau_{rx}$ is shear stress in the x-direction
perpendicular to the r-direction, which is given as,
\begin{eqnarray*}
{\tau}_{rx}&=&\mu_{nf}~\frac{\partial \tilde{v}_x}{\partial r} =\frac{\mu_{nf}}{\mu_{bf}}\left(\mu_{p}+\frac{{\tau}_{0}[1-exp(-m|\frac{\partial \tilde{v}_x}{\partial r}|)}{|\frac{\partial \tilde{v}_x}{\partial r}|}\right)\frac{\partial \tilde{v}_x}{\partial r} \\[3pt]
\end{eqnarray*}

\noindent \textbf{Energy equation:}

\begin{equation}
     (\rho c_p)_{nf}\left[\tilde{v}_x\frac{\partial \tilde{T}}{\partial x}+ \tilde{v}_r\frac{\partial \tilde{T}}{\partial r}\right]= k_{nf}\left[\frac{1}{r}\frac{\partial}{\partial r}\left(r\frac{\partial \tilde{T}}{\partial r}\right)\right]  \label{Eq:energy}
\end{equation}
Here, x and r denote the axial and radial coordinates, respectively. The velocity components in these directions are represented by 
\(\tilde v_x\) and \(\tilde v_r\), while $p$ is the fluid pressure. The effective density, viscosity, and thermal conductivity of the nanofluid are denoted by $\rho_{nf}$, $\mu_{nf}$, and $k_{nf}$, respectively, as summarized in  \autoref{tab:Thermo_models_agglomeration} for both aggregation and non-aggregation.

\subsection{\textbf{Boundary conditions}}
As shown in the \autoref{fig:Cylinder}, the conditions at the boundary of the domain are as follows:
\begin{equation}
\left.
\begin{aligned}
&\text{At } x = 0 \text{ and } 0 < r < r_1: \quad \tilde{v}_x = u_0,\quad \tilde{v}_r = 0,\quad  p = p_0,\quad \tilde{T}=T_0 .\\
&\text{At }  r = r_1   \text{ and at any value x } :\text{no-slip } \quad \tilde{v}_x = 0,\quad \tilde{v}_r = 0 \quad\text{and} \quad \tilde{T}=T_w.\\
&\text{At }  r=0  \quad\text{and at any value x}: \frac{\partial \tilde{v}_x}{\partial r} = 0,\quad \tilde{v}_r=0,\quad \frac{\partial \tilde{T}}{\partial r} = 0.\\
\end{aligned}
\right\}
\label{eq:BC_eqns}
\end{equation}
Using these hydrodynamic boundary conditions, the integral form of the continuity equation can be written as
\begin{eqnarray} 
\int_{0}^{r_1} 2\pi r \tilde{v}_x \, dr = \pi r_1^2 u_0.\label{eqn:integral_conty_eqn} 
\end{eqnarray}
\subsection{\textbf{Non-dimensionalization}}
In this section, we outline the parameters used to non-dimensionalize the governing equations and the boundary conditions. Given that the focus of this study is to analyze the effects of yield stress and volume fractions in single-phase homogeneous nanofluid models on heat transfer characteristics, the Reynolds number has been absorbed into some of the parameters as shown in \cite{baioumy2021bingham}.

The following are the parameters used to non-dimensionalize the governing equations and boundary conditions:
\begin{eqnarray}
 \begin{aligned}
&\eta= \frac{r}{r_1}, \quad  \xi= \frac{2x}{r_1 R_e},\quad U = \frac{\tilde{v}_x}{u_0}, \quad 
V = \frac{\tilde{v}_r}{u_0}\left(\frac{R_e}{2}\right), \quad P = \frac{p - p_0}{\rho_{bf} u_0^2}, \quad \Theta=\frac{\tilde{T} - T_w}{T_0- T_w}, \\
& 
\textit{M}= \frac{m u_0}{r_1}, \quad B_n = \frac{\tau_0 r_1}{u_0 \mu_b}, \quad 
R_e = \frac{2\rho_{bf} r_1 u_0}{ \mu_b}, 
\quad \textit{Pr}=\frac{\mu_b (c_p)_{bf}}{k_{bf}}\\
\end{aligned}
\label{eq:non_dim_parameters}
\end{eqnarray}

 \noindent Here, $B_n$ represents the Bingham number, \textit{M} denotes the regularization parameter, $R_e$ stands for the Reynolds number, and $Pr$ indicates the Prandtl number.

Using the non-dimensional parameters defined in Equation \eqref{eq:non_dim_parameters}, the governing equations can be expressed in their dimensionless form as follows:

\begin{eqnarray}
 \frac{\partial U}{\partial \xi}+\frac{1}{\eta}\frac{\partial (\eta V)}{\partial \eta}&=&0\label{Eq:ND_Continuity} \\
 U \frac{\partial U}{\partial \xi}+V\frac{\partial U}{\partial \eta}&=& -\frac{1}{\Phi_1}\frac{dP}{d \xi}+\left(\frac{1}{\Phi_2}\right)\frac{1}{ \eta}\frac{\partial}{\partial \eta}\left\{ \eta \left[1+\frac{ {B_n}\left(1-exp(\mathbf{-\textit{M}} |\frac {\partial U}{\partial \eta}| \right)}{|\frac {\partial U}{\partial \eta}|}\right]\frac {\partial U}{\partial \eta} \right\}  \label{Eq:ND_Momentum}\\
 \int_{0}^{1} U  \eta \, d \eta &= &\frac{1}{2} \label{Eq:ND_Intergral}\\\nonumber\\
 U \frac{\partial \Theta}{\partial \xi}+V\frac{\partial \Theta}{\partial \eta}&=&\left(\frac{\alpha_{nf}}{\alpha_{bf}}\right)\frac{1}{Pr}\left\{\frac{\partial^2 \Theta}{\partial \eta ^2} + \frac{1}{ \eta}\frac{\partial \Theta}{\partial \eta }\right \}\label{Eq:ND_Energy}
\end{eqnarray}
Where,
\begin{eqnarray*}
    \Phi_1&=&\frac{\rho_{nf}}{\rho_{bf}}
    \quad,\Phi_2=\left(\frac{\mu_{bf}}{\mu_{nf}}\right)\left(\frac{\rho_{nf}}{\rho_{bf}}\right) \quad,
    \left(\frac{\alpha_{nf}}{\alpha_{bf}}\right)=\left(\frac{k_{nf}}{k_{bf}}\right)\left(\frac{(\rho c_p)_{bf}}{(\rho c_{p})_{nf}}\right)
\end{eqnarray*}

and the corresponding dimensionless boundary conditions are as follows,
\begin{equation}
\left.
\begin{aligned}
&\text{At }  \xi = 0 \text{ and } 0 < \eta <1: \quad U = 1,\quad V = 0\quad  P=0,\quad \text{and} \quad \Theta=1 .\\
&\text{At } \eta=1  \text{ and at any value  $\xi$ } : \quad U = 0,\quad V= 0 \quad\text{and} \quad \Theta=0.\\
&\text{At } \eta=0  \quad\text{and at any value  $\xi$ }: \frac{\partial U}{\partial \eta} = 0,\quad V=0,\quad \frac{\partial \Theta}{\partial \eta} = 0\\
\end{aligned}
\right\}
\label{eq:ND_BC_eqns}
\end{equation}
The above set of equations \eqref{Eq:ND_Continuity}, \eqref{Eq:ND_Momentum}, \eqref{Eq:ND_Intergral}, \eqref{Eq:ND_Energy} is used to calculate the hydrodynamic and thermal properties of the flow domain.

\subsection{\textbf{Limitations of the single-phase nanofluid models}}
It should be noted that single-phase nanofluid viscosity models, such as the Brinkman and Krieger–Daugherty models, assume the nanofluid behaves as a homogeneous continuum, despite the presence of dispersed nanoparticles. The primary parameter which depicts this is the
volume fraction. Additionally, thermal and velocity equilibrium between the nanoparticles and base fluid is assumed, allowing properties such as density, velocity, thermal conductivity, and specific heat to be treated as effective, volume-fraction-dependent, and temperature-independent quantities, each with a limited range of validity.

However, these models neglect important physical phenomena, including Brownian motion, thermophoresis, and gravitational settling. Even in aggregation-based models, particle clustering is simplified using an aggregate-to-primary particle size ratio, without accounting for shear-induced breakup or migration near walls. A more accurate description of such effects requires full-scale simulations of the Navier–Stokes equations coupled with particle dynamics and fluid–particle interactions.

\subsection{\textbf{Friction coefficient, and Nusselt number:}}

The friction coefficient, $C_f$, is defined as follows:
\begin{equation}
    C_f = \frac{2\tau_w}{\rho_{bf} u_0^2}, 
    \qquad 
    \tau_w = \mu_{nf} \left. \frac{\partial \tilde{v}_x}{\partial r} \right|_{r = r_1}
    \label{eq:Cf}
\end{equation}

The local Nusselt number is expressed as:
\begin{equation}
    Nu = \frac{2r_1 q_w}{k_{bf}(T_w - T_b)}, 
    \qquad 
    q_w = k_{nf} \left. \frac{\partial \tilde{T}}{\partial r} \right|_{r = r_1}
    \label{eq:Nu}
\end{equation}

where $T_w$ is the wall temperature and  ${T}_b$ is the bulk temperature in the domain given by
\begin{equation}
{T_b} = \frac{\int_0^{r_1} 2\pi r \tilde{v}_x \tilde{T} \, dr}{\int_0^{r_1} 2\pi r \tilde{v}_x \, dr} 
\end{equation}
After non-dimensionalisation, the friction coefficient, Nusselt number, and bulk temperature are given as follows,

\begin{eqnarray}
    C_f &=& \frac{4}{R_e}\frac{\mu_{nf}}{\mu_{bf}}\left[1+\frac{B_n [1-exp(-\textit{M}|\frac{\partial U}{\partial \eta}|)]}{|\frac{\partial U}{\partial \eta}|}\right]\frac{\partial U}{\partial \eta} 
    \label{eq:Non_dim_Cf}\\\nonumber\\
    Nu & = &\frac{-2 \left. \frac{\partial \Theta}{\partial \eta} \right|_{w}\left(\frac{k_{nf}}{k_{bf}}\right)}{ \Theta_b}  \label{eq:Non_dim_Nu} \\\nonumber\\
        \Theta_b&=&2\int_o^{\eta}  U \eta \Theta \,d\eta \label{eq:Non-dim Tb}
\end{eqnarray}
\section{Computational Scheme}
The equations \eqref{Eq:ND_Continuity}–\eqref{Eq:ND_Energy} are nonlinear, which makes obtaining analytical solutions difficult. The nonlinear behavior results from the presence of coupled terms, and attempting to simplify them often leads to a loss of accuracy. Hence, numerical methods are generally used to solve such equations efficiently.

\begin{figure}[H]
    \centering
    \includegraphics[width=0.48\linewidth]{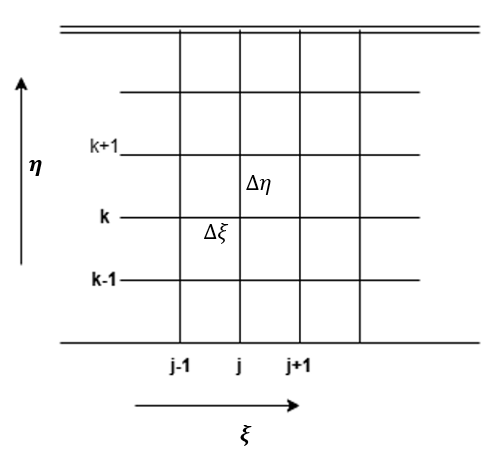}
    \caption{Finite Difference Scheme}
    \label{fig:grid_gen}
\end{figure}
The Finite Difference Method (FDM), as adopted from \cite{venthan2021theoretical}, incorporates the necessary boundary conditions within its numerical formulation. In the present study, a central difference scheme is employed in the radial direction, while a backward difference scheme is applied in the axial direction. The grid spacings are denoted by 
\(\Delta \eta\) in the radial direction and 
\(\Delta \xi\)  in the axial direction.

\autoref{fig:grid_gen} presents the grid generation procedure along the axial direction. At a typical grid point 
\((j,k)\), the flow variables are assumed to be known, whereas the values at the subsequent axial location 
\((j+1,k)\) are unknown and must be determined.
\begin{equation}
\left.
\begin{aligned}
& U\frac{\partial U}{\partial \xi}=U_{j,k}\left[\frac{U_{j+1,k}-U_{j,k}}{\Delta \xi}\right]\\
   & V\frac{\partial U}{\partial \eta}=V_{j,k} \left[\frac{U_{j+1,k+1}-U_{j+1,k-1}}{2\Delta \eta}\right]\\[2pt]
   & \frac{ dP}{d\xi}=\frac{P_{j+1}-P_j}{\Delta \xi}\\
   & \frac{1}{\eta}\frac{\partial U}{\partial \eta}=\frac{1}{\eta_k}\frac{U_{j+1,k+1}-U_{j+1,k-1}}{2\Delta \eta}\\[2pt]
   &\frac{\partial^2 U}{\partial \eta^2}=\frac{U_{j+1,k+1}-2U_{j+1,k}+U_{j+1,k-1}}{(\Delta \eta)^2}\\[2pt]
   &\frac{\partial U}{\partial \xi}=\frac{U_{j+1,k+1}-U_{j,k+1}+U_{j+1,k}-U_{j,k}}{2\,\Delta \xi}\\[2pt]
   &\frac{1}{\eta} \frac{\partial( V\eta)}{\partial \eta}=\frac{1}{\eta_k}
\frac{V_{j+1,k+1} \eta_{k+1}-V_{j+1,k} \eta_k}{\Delta \eta}
\end{aligned}
\right\}
\end{equation}
The discretization approaches employed for the continuity \eqref{Eq:ND_Continuity}, momentum \eqref{Eq:ND_Momentum}, and integral equation \eqref{Eq:ND_Intergral} equations are as follows:

\begin{align}
&\frac{U_{j+1,k+1}-U_{j,k+1}+U_{j+1,k}-U_{j,k}}{2\,\Delta \xi}
+ \frac{1}{\eta_k}
\frac{V_{j+1,k+1}\eta_{k+1}-V_{j+1,k}\eta_k}{\Delta \eta}
= 0
 \\[6pt]
&\text{which simplifies to}
\nonumber \\[6pt]
&V_{j+1,k+1}
=
\left(\frac{\eta_k}{\eta_{k+1}}\right)
V_{j+1,k}
-
\frac{\Delta \eta}{2\,\Delta \xi}
\left(\frac{\eta_k}{\eta_{k+1}}\right)
\left[
U_{j+1,k+1}-U_{j,k+1}
+
U_{j+1,k}-U_{j,k}
\right]
\label{Eq:FDM_cont}\\\nonumber\\[6pt]
&U_{j.k}\frac{U_{j+1,k}-U_{j,k}}{\Delta\xi}+V_{j,k}\frac{U_{j+1,k+1}-U_{j+1,k-1}}{2\Delta\eta}=-\frac{1}{\Phi_1}\frac{P_{j+1}-P_{j}}{\Delta\xi}\nonumber\\[3pt]
&~~~~~~~~~~~~~~~~~~~~~~~+\frac{B_1}{\eta_k\Phi_2}\frac{U_{j+1,k+1}-U_{j+1,k-1}}{2\Delta\eta}+\frac{B_2}{\Phi_2}\frac{U_{j+1,k+1}-2U_{j+1,k}+U_{j+1,k-1}}{(\Delta \eta)^2}
\end{align}
\begin{align}
&\text{which simplifies to}
\nonumber \\[6pt]
& P_{j+1}\left(\frac{1}{\Phi_1}\right)+ U_{j+1,k-1}
\left[
\frac{\Delta \xi \left(\frac{B_1}{\Phi_2}\right)}{2 \Delta \eta \, \eta_k}
- \frac{\Delta \xi \, V_{j,k}}{2 \Delta \eta}
- \frac{\Delta \xi \left(\frac{B_2}{\Phi_2}\right)}{(\Delta \eta)^2}
\right] \nonumber \\[3pt]
&+ U_{j+1,k}
\left[
U_{j,k}
+ \frac{2 \Delta \xi \left(\frac{B_2}{\Phi_2}\right)}{(\Delta \eta)^2}
\right] + U_{j+1,k+1}
\left[
-\frac{\Delta \xi \left(\frac{B_1}{\Phi_2}\right)}{2 \Delta \eta \, \eta_k}
+ \frac{\Delta \xi \, V_{j,k}}{2 \Delta \eta}
- \frac{\Delta \xi \left(\frac{B_2}{\Phi_2}\right)}{(\Delta \eta)^2}
\right] \nonumber \\[3pt]
&= (U_{j,k})^2
+ P_j\left(\frac{1}{\Phi_1}\right)
\label{Eq:FDM_momen}\\[6pt]
&\text{here, }B_1 = 1 + \frac{B_n \left[1 - \exp\left(-M \left|\frac{\partial U}{\partial \eta}\right|\right)\right]}
{\left|\frac{\partial U}{\partial \eta}\right|}, \quad
B_2 = 1 + B_n M \exp\left(-M \left|\frac{\partial U}{\partial \eta}\right|\right),\nonumber\\
&\Phi_1=\frac{\rho_{nf}}{\rho_{bf}},
\quad\Phi_2=\left(\frac{\mu_{bf}}{\mu_{nf}}\right)\left(\frac{\rho_{nf}}{\rho_{bf}}\right) \quad \nonumber
\end{align}

The integral equation of continuity\eqref{Eq:ND_Intergral} in finite difference form is
\begin{align}
\sum_{k=0}^{n} \eta_{k}U_{j+1,k}=\sum_{k=0}^{n} \eta_{k}U_{j,k}.
\label{Eq:FDM_int}  
\end{align}

To determine the unknowns U, V, and P in the equations \eqref{Eq:FDM_cont}, \eqref{Eq:FDM_momen}, and \eqref{Eq:FDM_int}, the terms with the subscript (j) are considered to be known, while those with (j+1) are treated as unknowns. The linearized algebraic equations  \eqref{Eq:FDM_momen} and \eqref{Eq:FDM_int} are solved numerically to obtain \(U_{j+1,k}  \text{and }  P_{j+1}\). These updated values of U are then substituted into equation \eqref{Eq:FDM_cont} to compute \(V_{j+1,k}\). This procedure is repeated successively in the downstream direction to obtain U, P, and V throughout the flow domain for the nanofluid.

To address the thermal aspect of the domain, we have employed the energy equation \eqref{Eq:ND_Energy} in its finite difference form as given below.

\begin{align}
 U_{j,k}\frac{\Theta _{j+1,k}-\Theta_{j,k}}{\Delta \xi}&+V_{j,k}\frac{\Theta_{j+1,k+1}-\Theta_{j+1,k-1}}{2\Delta \eta}\nonumber\\=&\left(\frac{\alpha_{nf}}{\alpha_{bf}}\right)\frac{1}{Pr}\left[\frac{ \Theta_{j+1,k+1}-2\Theta_{j+1,k}+\Theta_{j+1,k-1}}{(\Delta \eta)^2}+\frac{1}{\eta_k}\frac{\Theta_{j+1,k+1}-\Theta_{j+1,k-1}}{2\Delta \eta}\right].
 \label{Eq:FDM_ENG}  
\end{align}

For $\eta=0$,  the finite difference representation cannot be directly applied. Therefore, the limit of equation \eqref{Eq:ND_Energy} as 
$\eta$ approaches 0 is first evaluated, and an equivalent finite difference formulation is then derived based on this limiting form.

i.e.
\text{ for $\eta$=0}
\begin{align}
    U_{0,k}\frac{\Theta_{j+1,0}-\Theta_{j,0}}{\Delta \xi}=\left(\frac{\alpha_{nf}}{\alpha_{bf}}\right)\frac{4}{Pr}\frac{\Theta_{j+1,1}-\Theta_{j+1,0}}{(\Delta \eta)^2}\label{Eq:FDM_ENG_R_0}
\end{align}
The Equations \eqref{Eq:FDM_ENG}  and \eqref{Eq:FDM_ENG_R_0} simplify to:

\begin{align}
&\Theta_{j+1,k-1} \left[- \frac{ V_{j,k}}{2 (\Delta \eta)}+\left(\frac{\alpha_{nf}}{\alpha_{bf}}\right)\frac{1}{2Pr (\Delta \eta) \eta_{k}}  - \left(\frac{\alpha_{nf}}{\alpha_{bf}}\right)\frac{1}{Pr(\Delta \eta)^2} \right]\nonumber\\[3pt]
&+\Theta_{j+1,k} \left[ \frac{U_{j,k}}{\Delta \xi} + \left(\frac{\alpha_{nf}}{\alpha_{bf}}\right)\frac{2}{Pr(\Delta \eta)^2} \right] \nonumber\\[3pt]
&+ \Theta_{j+1,k+1} \left[-\left(\frac{\alpha_{nf}}{\alpha_{bf}}\right)\frac{1}{2Pr (\Delta \eta) \eta_{k}} + \frac{ V_{j,k}}{2 (\Delta \eta)} - \left(\frac{\alpha_{nf}}{\alpha_{bf}}\right)\frac{1}{Pr(\Delta \eta)^2} \right]= \frac{U_{j,k} \Theta_{j,k}}{\Delta \xi}\label{Eq:FDM_Eng1}   \\[3pt]
&\text{and}\nonumber\\
&\Theta_{j+1,0} \left[ \frac{U_{j,0}}{\Delta \xi} +\left(\frac{\alpha_{nf}}{\alpha_{bf}}\right) \frac{4}{Pr(\Delta \eta)^2} \right]+ \Theta_{j+1,1} \left[\left(\frac{\alpha_{nf}}{\alpha_{bf}}\right)\frac{-4}{Pr (\Delta \eta)^2}\right]= \frac{U_{j,0} \Theta_{j,0}}{\Delta \xi}\label{Eq:FDM_Eng2}  \end{align}

Equations \eqref{Eq:FDM_Eng1} and \eqref{Eq:FDM_Eng2} are treated as linearized algebraic relations. By applying a suitable numerical scheme, the unknown values of temperature at the downstream location  \((j+1,k)\) can be evaluated. Repeating this procedure step by step enables the computation of the temperature field.
Once the temperature distribution within the domain is obtained, it is further utilised to calculate heat transfer characteristics, including the Nusselt number and bulk temperature. This framework facilitates the investigation of nanoparticle effects in a viscoplastic base fluid, here Bingham fluid.

\section{Grid Independence Study and Validation}
In this section, we outline a step-by-step procedure to conduct a grid independence study. The following subsection provides details on the Bingham-Papanastasiou model and the selection of grid size and validation of our results.

\subsection{Rheogram of Bingham-Papanastasiou Model }
\begin{figure}[htbp]
    \centering
    \includegraphics[width=0.5\linewidth]{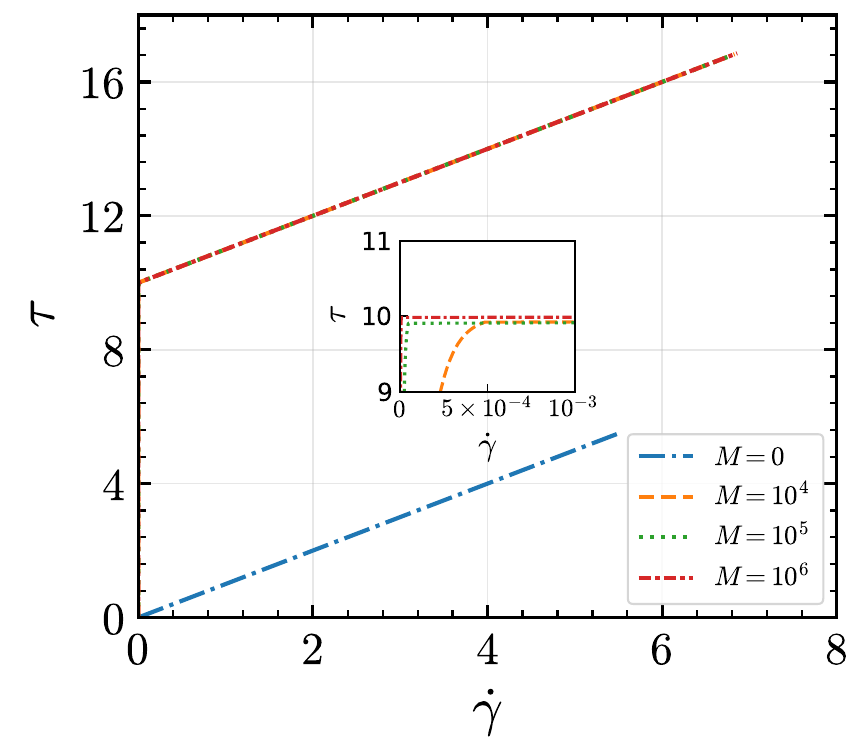}
    \caption{Rheogram for the Bingham-Papanastasiou Model \cite{article} showing the trends for different values of the regularization parameter, $M$}
    \label{fig:M_Variation}
\end{figure}

In our analysis, we have evaluated the yield stress behaviour using 16,601 grids in the axial direction and 501 grids in the radial direction for different values of the regularization parameter, $M$. \autoref{fig:M_Variation} illustrates the relationship between shear stress($\tau$) and shear rate($\dot\gamma$) for a Bingham number, $B_n = 10$. Our models indicate that when \(M=0\), the behaviour resembles that of a Newtonian fluid. As the value of M increases, the behaviour approaches the ideal Bingham viscoplastic model. In our simulations, the value of \(M= 10^{6}\) is sufficient to represent the ideal Bingham viscoplastic model for different yield stresses (i.e., Bingham numbers).

\subsection{Choice of the Grid size}
\begin{figure}[htbp]
    \centering
    \begin{subfigure}{0.48\textwidth}
        \centering
        \includegraphics[width=\textwidth]{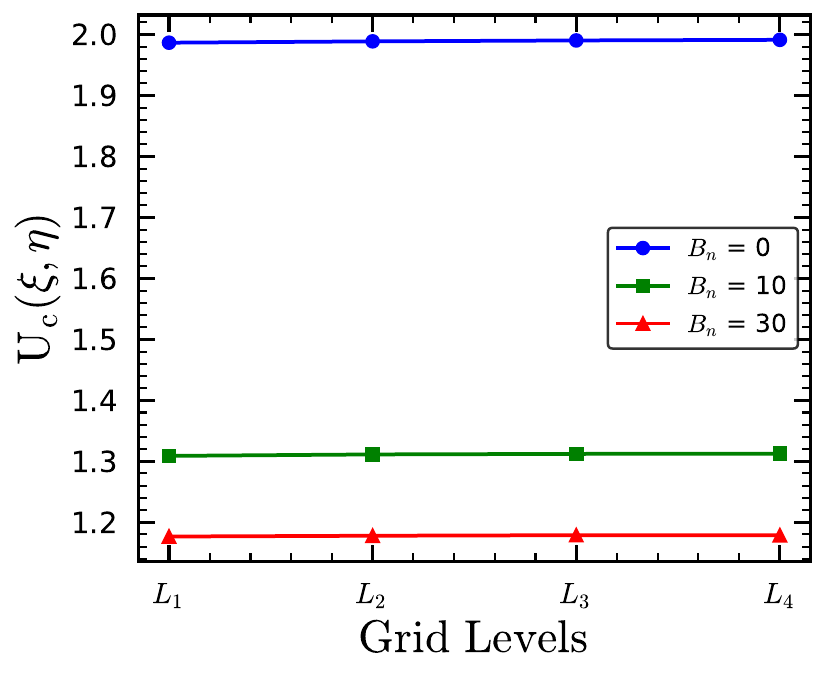}
        \caption{Variation of Centerline axial velocity with Grid levels for different values of Bingham number, $B_n$.}
        \label{fig:Grid_B}
    \end{subfigure}
     \hfill
    \begin{subfigure}{0.48\textwidth}
        \centering
        \includegraphics[width=\textwidth]{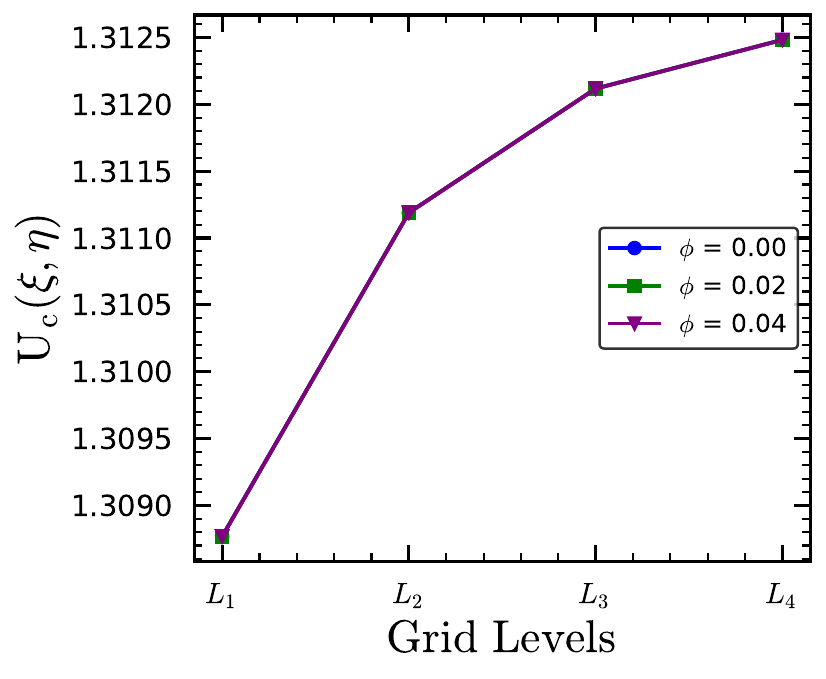}
        \caption{Variation of Centerline axial velocity with Grid levels for different values of Volume Fraction, $\phi$.}
        \label{fig:Grid_phi}
    \end{subfigure}
\caption{Grid independence study for different grid resolutions, $L_1$:($8471 \times 256$), $L_2$:($11857 \times 358$),  \\ $L_3$ :($16601 \times 501$), and $L_4$:($23215 \times 667$).}
    \label{fig:Grid_GCI}
\end{figure}

\begin{figure}[htbp]
    \centering
    \begin{subfigure}[b]{0.48\linewidth}
        \centering
        \includegraphics[width=\linewidth]{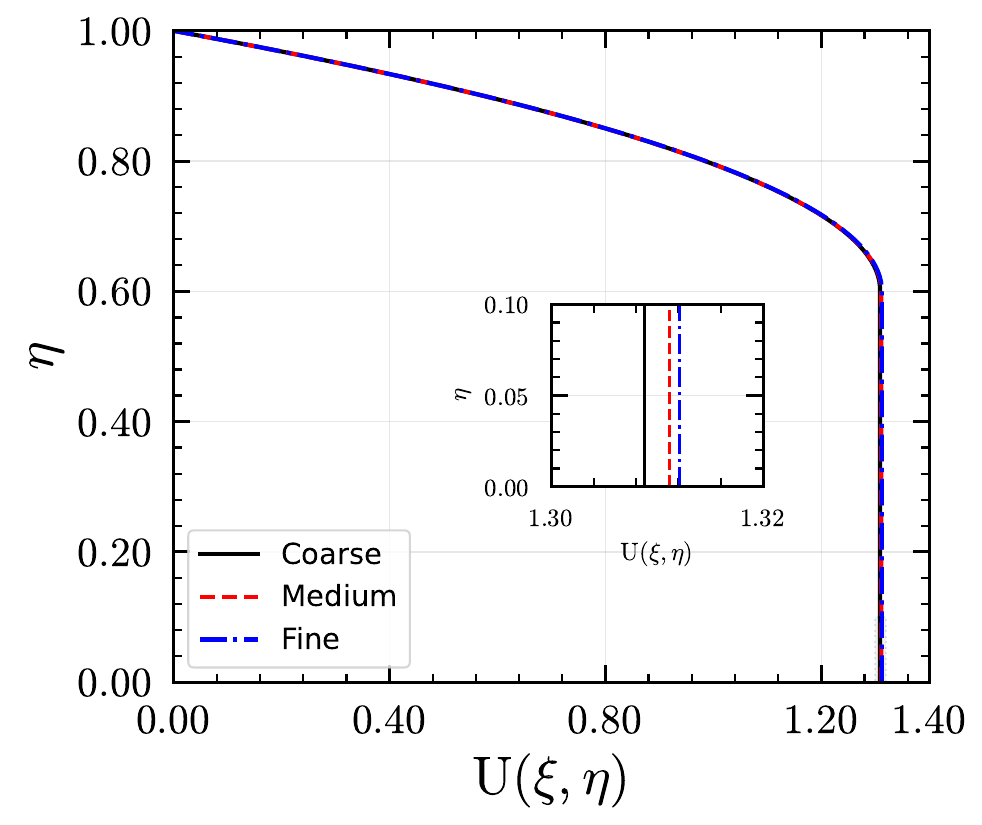}
        \caption{Axial velocity }
        \label{Fig:Grid_study_Velocity}
    \end{subfigure}
      \hfill
        \begin{subfigure}[b]{0.48\linewidth}
        \centering
        \includegraphics[width=\linewidth]{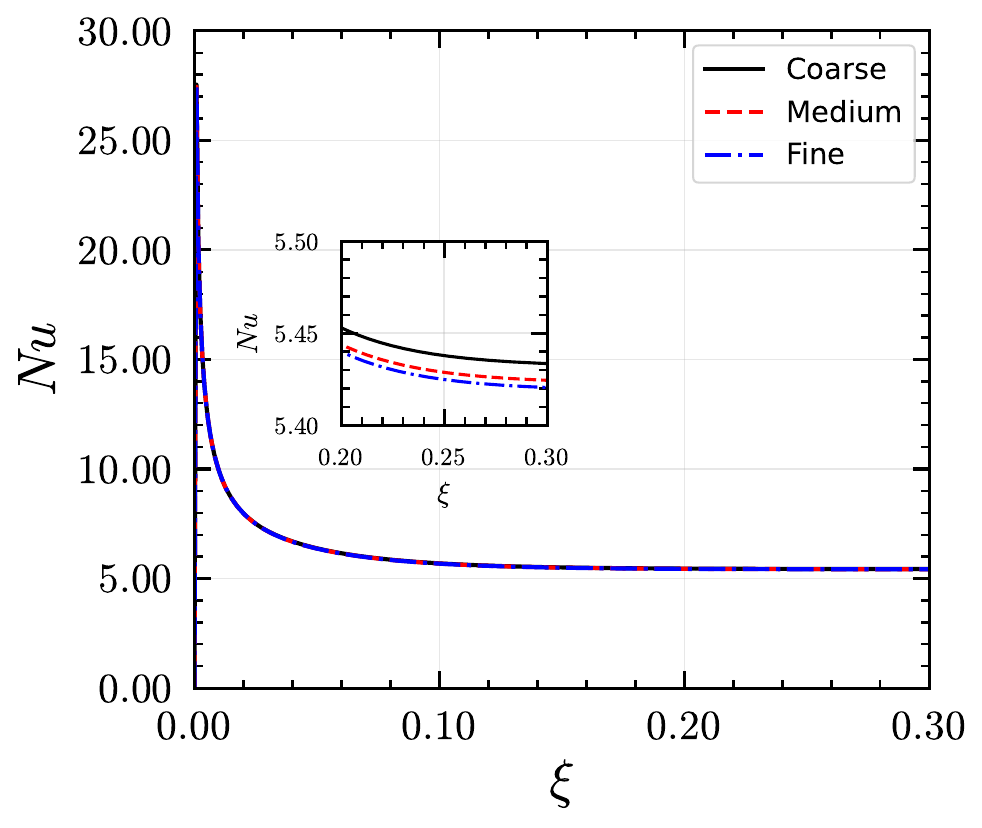}
        \caption{Nusselt Number}
        \label{Fig:Grid_study_Nusselt}
    \end{subfigure}
    \caption{Grid independent Study for axial velocity ($U(\xi,\eta)$ and Nusselt number($Nu$) with Coarse grid $L_1$: ($8471 \times 256$), Medium grid $L_2$: ($11857 \times 358$), and Fine grid $L_3$: ($16601 \times 501$). }
    \label{fig:Grid_Vel_Nu}
\end{figure}

We fix the value of $M=10^6$ and conduct the grid independence study for the desired range of $Bn$, which is $[0,30]$ as shown in the \autoref{fig:Grid_B}. Subsequently, the range of volume fraction, $\phi$, which is $[0,5\%]$, is also examined using the same grid configuration, as shown in  \autoref{fig:Grid_phi}.
 The results presented in \autoref{fig:Grid_B} and \autoref{fig:Grid_phi} collectively validate the selection of 16,601 nodes in the axial direction and 501 nodes in the radial direction, ensuring adequate resolution for accurate computations. The grid independence analysis has been carried out for both axial velocity and Nusselt number, as depicted in \autoref{fig:Grid_Vel_Nu}, confirming that the fine grid of 
$16601 \times 501$ is sufficiently refined for reliable results (For GCI of our results, ref \autoref{Sec:GCI}).

\subsection{Validation}
\begin{figure}[htbp]
    \centering
        \includegraphics[width=0.5\linewidth]{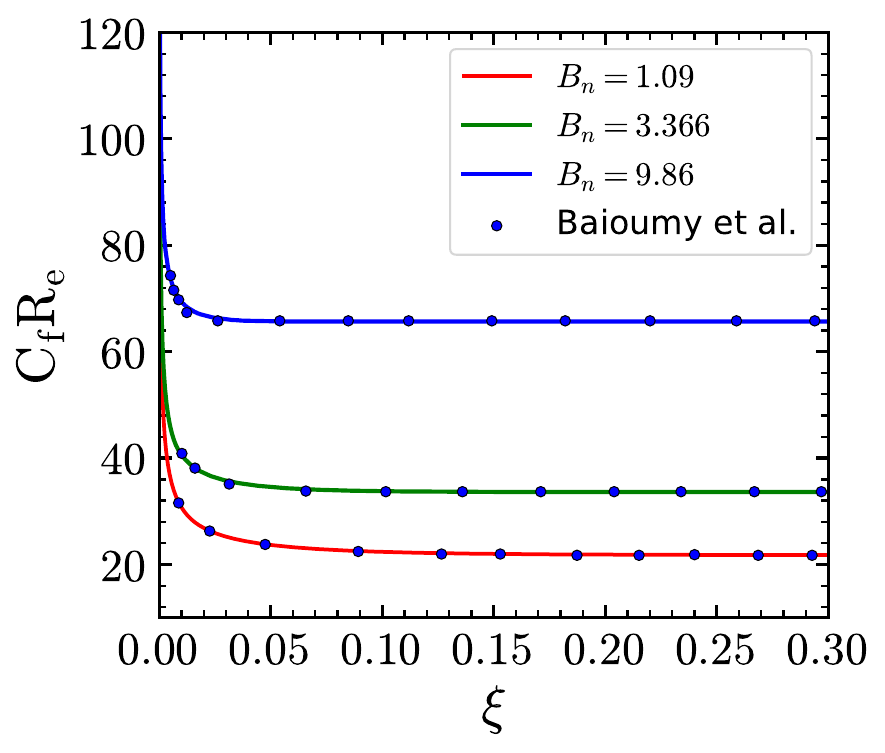}
    \caption{Comparison of friction factor at the wall, $C_f R_e$, along the axial direction for different Bingham numbers compared with Baioumy \textit{et al} \cite{baioumy2021bingham}.}
    \label{fig:Val_CF}
\end{figure}

\begin{figure}[htbp]
    \centering
    \includegraphics[width=0.5\linewidth]{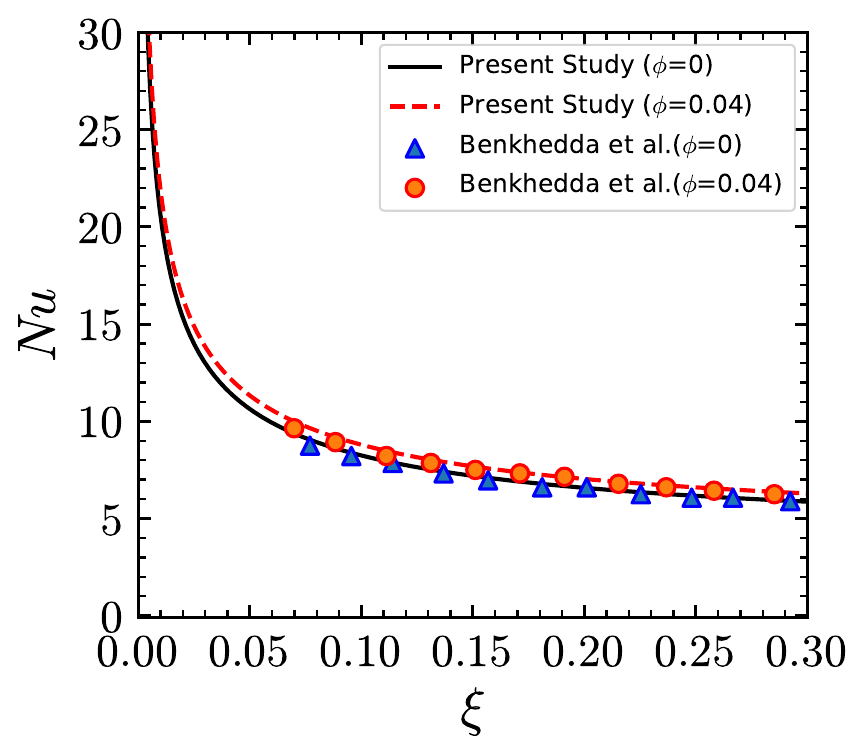}
    \caption{Comparison of Nusselt number, $Nu$, along the axial direction for different volume fractions and Pr=6.2 with constant heat flux boundary condition compared with Benkhedda \textit{et al} \cite{benkhedda2020convective}.}
    \label{fig:Val_NU}
\end{figure}

In this section we present the validation of our results with the earlier studies.
 \autoref{fig:Val_CF} shows the variation of friction factor at the wall ($C_f R_e$), along the non-dimensional axial direction $\xi$ for different values of Bingham number and compared with Baioumy \textit{et al} \cite{baioumy2021bingham}. The predicted friction coefficient shows good agreement with the data shown by Baioumy et al. 
 \cite{baioumy2021bingham} for $B_n= 1.09, 3.36, 9.86$ with a maximum deviation of $1.27\%$(ref \autoref{Sec_Validation}) near the inlet for $B_n=9.86$. Further, \autoref{fig:Val_NU} shows the variation of the Nusselt number, $Nu$, along $\xi$ for different values of the volume fraction $\phi$ and compared with the results of Benkhedda \textit{et al} \cite{benkhedda2020convective} for $\phi=0, \;0.04$. This Figure shows the present results match well with Benkhedda \textit{et al} \cite{benkhedda2020convective} with a maximum deviation of $3.62\%$ (ref \autoref{Sec_Validation}) in the developing region for $\phi = 0.04$.

\section{Results and discussion}
An investigation of a viscoplastic nanofluid flow in the entrance region of a circular cylinder has been conducted for the case of both non-aggregated and aggregated nanoparticles. The dimensionless governing equations are based on the Prandtl boundary layer theory, and the viscoplastic nanofluid is modeled using the Bingham-Papanastasiou approach. In this study, we compared both cases—with and without aggregation—using different models. For the non-aggregation case, the Brinkman viscosity model and Maxwell thermal conductivity model were employed. For the aggregation case, the Krieger-Dougherty viscosity model and Maxwell-Bruggeman thermal conductivity model were used. The governing equations, including the axial momentum, continuity, and energy equations, along with the integral form of the continuity equation, were solved numerically via a linearized finite difference method. The obtained solution reduces to the Newtonian case when the Bingham number is set to zero. Key flow characteristics such as the axial velocity (U), pressure drop (P), friction factor ($C_f$), bulk temperature ($\Theta_b$), Nusselt number (Nu), and performance evaluation criterion (PEC) have been analyzed and discussed. The parameters considered in this study include the Bingham number ($0 \leq B_n \leq 30$), the nanoparticle volume fraction ($0 \leq \phi \leq 0.05$) and $Pr=1$. We have analysed both the effect of volume fractions and the yield stress on the flow parameters in the subsections, namely, \autoref{Sec:Effect_of_Phi} and \autoref{Sec:Effect_of_Bn} .

\subsection{\textbf{Effect of volume fraction of nanoparticle in viscoplastic fluid:}}
\label{Sec:Effect_of_Phi}

In this section, we analyze the effect of adding nanoparticles to a viscoplastic fluid characterized by a Bingham number, $B_n = 10$, with nanoparticle volume fractions ranging from 0 to 0.05 for both non-aggregation and aggregation cases.

\begin{figure}[htbp]
    \centering
    \begin{subfigure}[b]{0.48\linewidth}
        \centering
        \includegraphics[width=\linewidth]{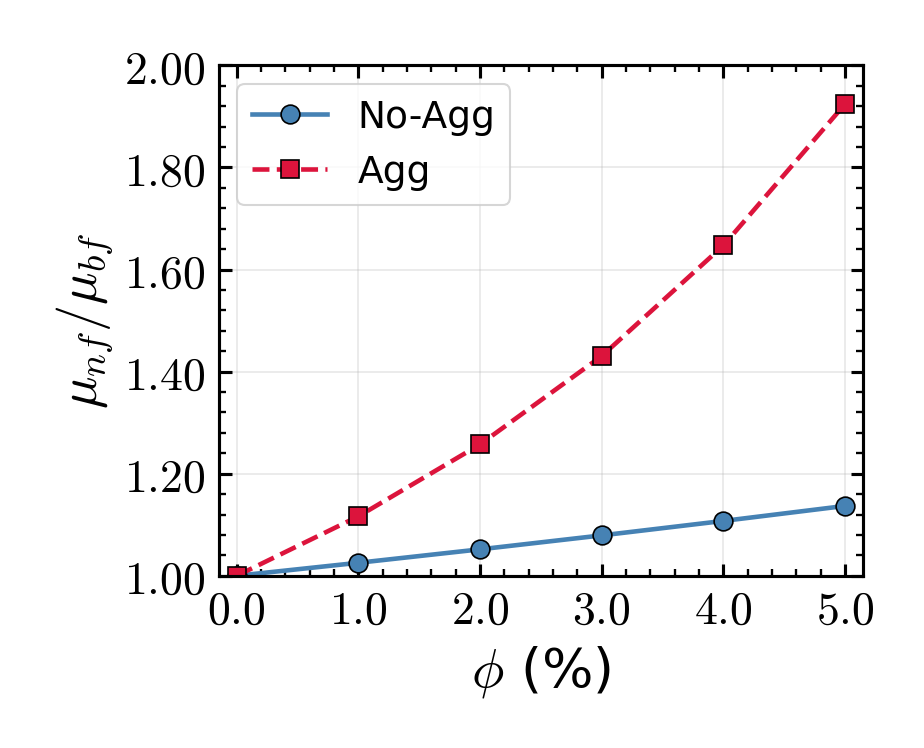}
        \caption{Effective viscosity }
        \label{Fig:Viscosity_ratio}
    \end{subfigure}
      \hfill
        \begin{subfigure}[b]{0.48\linewidth}
        \centering
        \includegraphics[width=\linewidth]{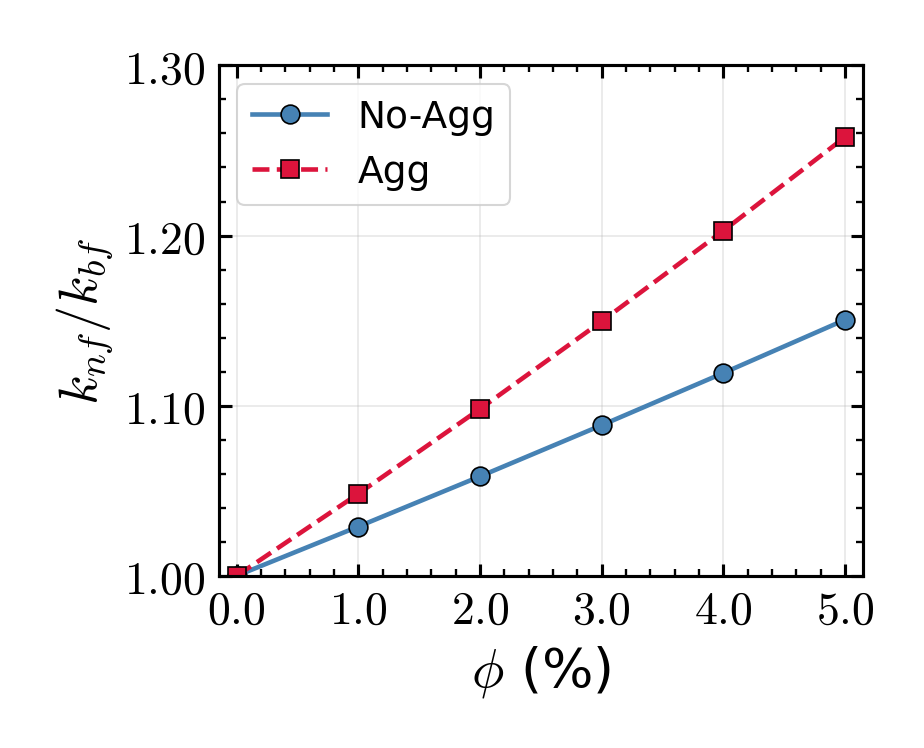}
        \caption{Effective thermal conductivity}
        \label{Fig:Conductivity_ratio}
    \end{subfigure}
    \caption{Effective thermo-physical properties for both non-aggregation and aggregation models for different volume fractions of nanofluids.}
\end{figure}

 \autoref{Fig:Viscosity_ratio} and \autoref{Fig:Conductivity_ratio} demonstrate the effects of non-aggregation and aggregation on thermophysical properties such as effective viscosity and effective thermal conductivity as the volume fractions increase. The results indicate that aggregated nanoparticles enhance these properties compared to non-aggregated nanoparticles.

\begin{figure}[htbp]
    \centering
    \includegraphics[width=0.5\linewidth]{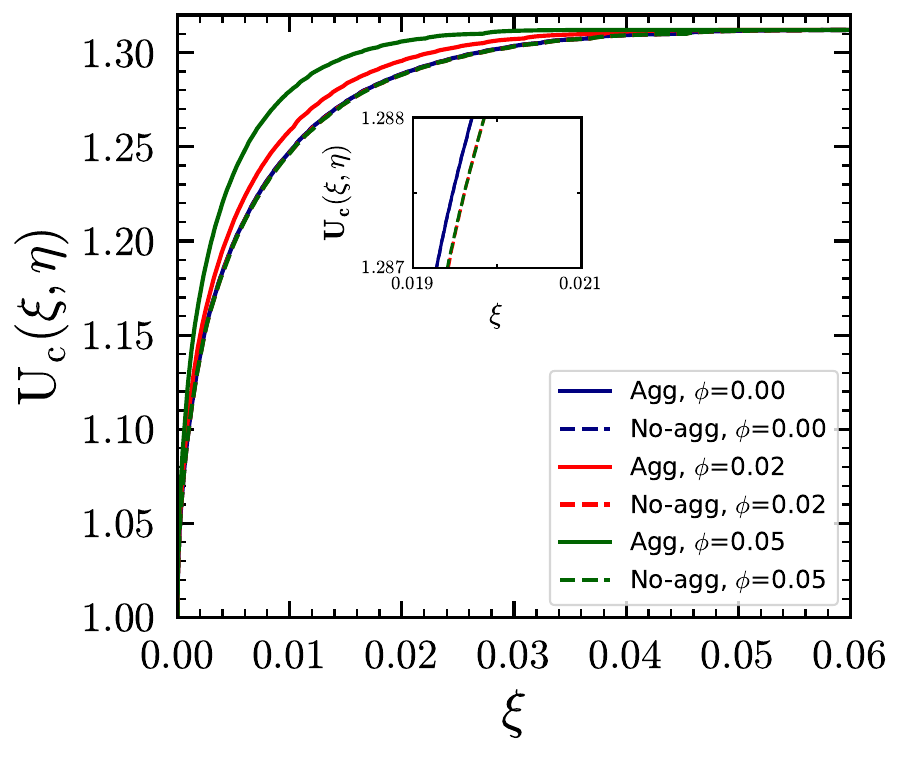}
    \caption{Centerline velocity in axial direction for different volume fractions $\phi=0,0.02,0.05$ and $B_n=10$}
    \label{fig:Center_Vel_phi}
\end{figure}

The effects of aggregation and non-aggregation on centerline axial velocity development within the cylinder are shown in \autoref{fig:Center_Vel_phi} for different volume fractions. In the non-aggregation case, there is minimal change in velocity in the developing region, and the velocity boundary layer develops similarly to that of the base fluid across all volume fractions. However, in the aggregation case, the velocity develops more rapidly than in the non-aggregation scenario.

\begin{figure}[htbp]
    \centering
    \begin{subfigure}[b]{0.48\linewidth}
        \centering
        \includegraphics[width=\linewidth]{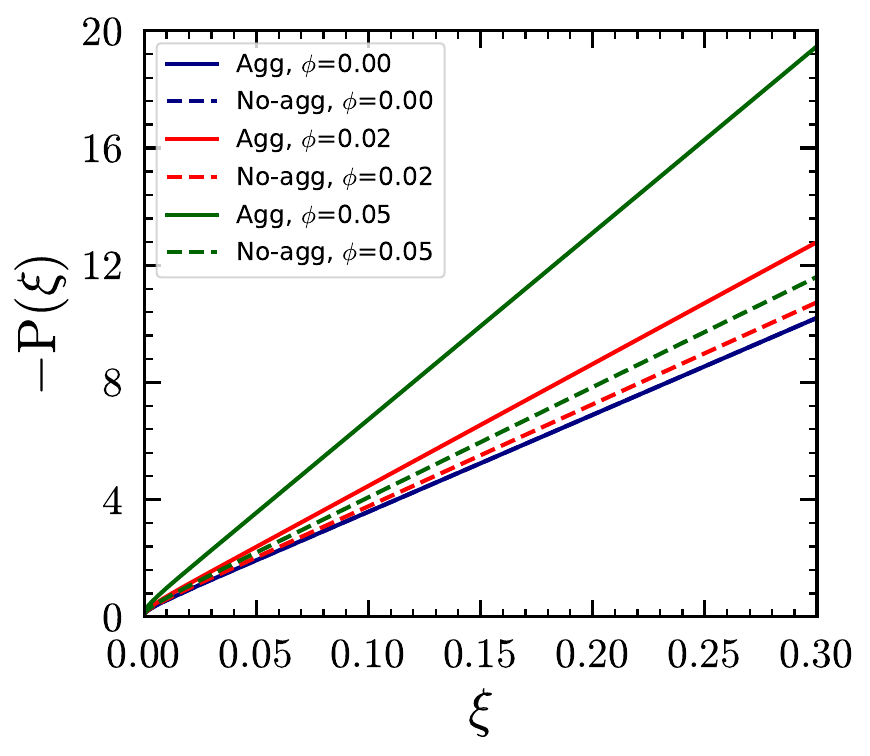}
        \caption{Pressure drop}
        \label{fig:P_B_10}
    \end{subfigure}
      \hfill
        \begin{subfigure}[b]{0.48\linewidth}
        \centering
        \includegraphics[width=\linewidth]{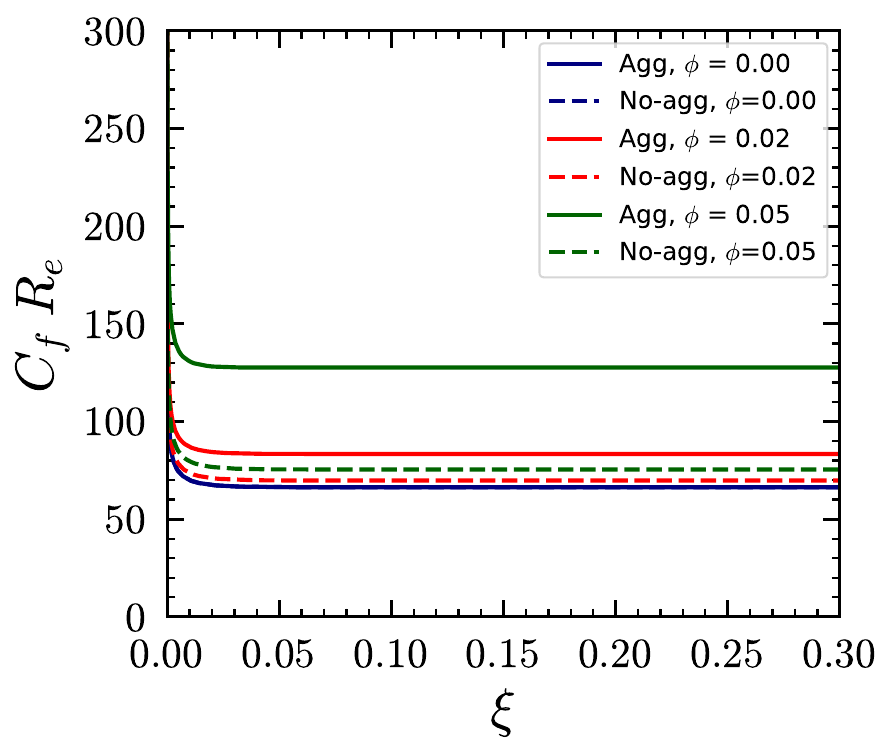}
        \caption{Friction coefficient variation}
        \label{fig:Cf_B_10}
    \end{subfigure}
    \vspace{0.3cm}

    \begin{subfigure}[b]{0.48\linewidth}
        \centering
        \includegraphics[width=\linewidth]{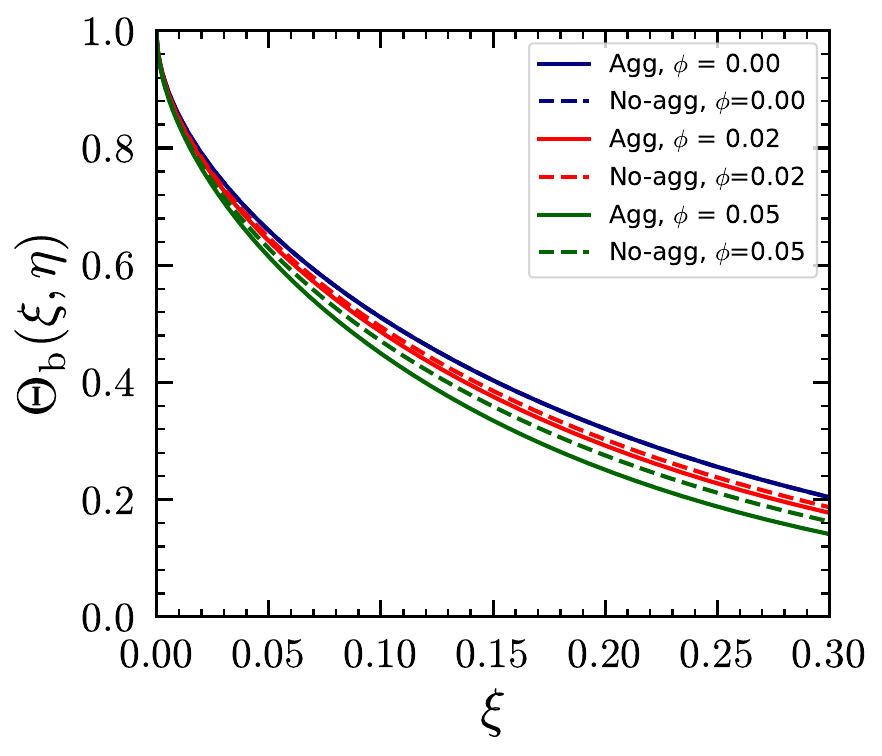}
        \caption{Bulk temperature}
        \label{fig:Tb_B_10}
    \end{subfigure}
    \hfill
    \begin{subfigure}[b]{0.48\linewidth}
        \centering
        \includegraphics[width=\linewidth]{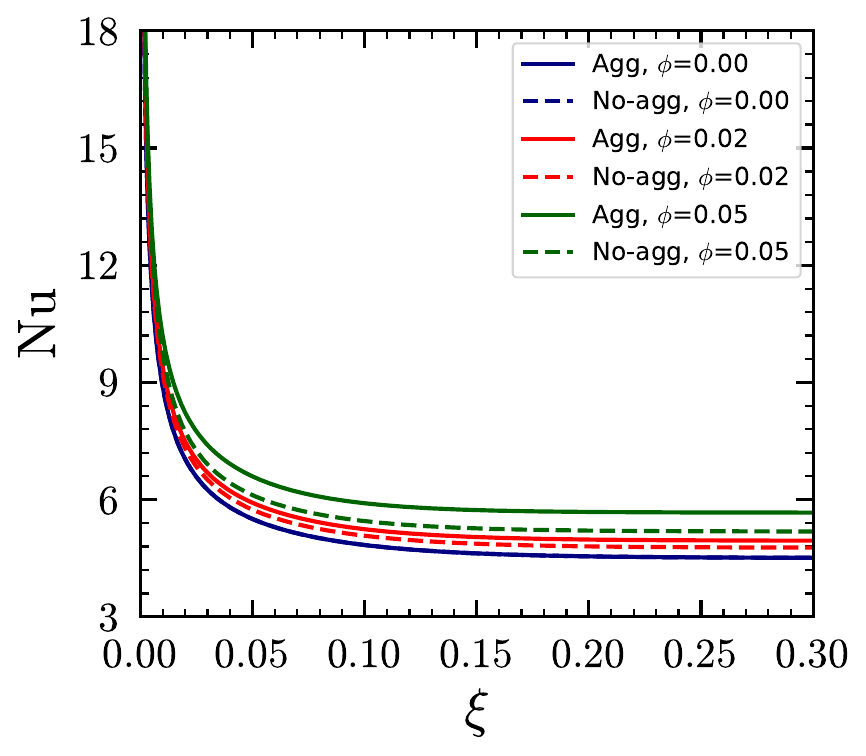}
        \caption{Nusselt number variation}
        \label{fig:Nu_B_10}
    \end{subfigure}

    \caption{Variation of Pressure drop (\autoref{fig:P_B_10}), Friction coefficient (\autoref{fig:Cf_B_10}), Bulk temperature (\autoref{fig:Tb_B_10}) and Nusselt number (\autoref{fig:Nu_B_10}) along the axial direction for different values of Volume fraction considering both non-aggregation and aggregation models with Bingham number $B_n=10$.}
    \label{fig:combined_B=10}
\end{figure}

Additionally, a higher volume fraction indicates an earlier development of the boundary layer due to a rise in the effective viscosity which in turn increases the pressure drop as illustrated in \autoref{fig:P_B_10}. The pressure drop varies linearly along the axial direction, and a similar trend is observed after the addition of nanoparticles to the Bingham fluid, although with a higher pressure drop at larger volume fractions.


The rise in nanoparticle volume fraction results in a higher effective viscosity of the fluid. This enhancement in viscosity intensifies the wall shear stress, which in turn influences the friction coefficient at the wall, as described by Eq.\eqref{eq:Non_dim_Cf}. \autoref{fig:Cf_B_10}  depicts the axial variation of the wall friction coefficient, showing a gradual decline along the flow direction. As the effective viscosity becomes greater with higher $\phi$ , larger nanoparticle concentrations correspond to elevated friction factors. In the non-aggregation case, the effective viscosity remains nearly unchanged and close to unity; hence, the friction coefficient exhibits only minor variation. In contrast, under aggregation conditions, a sharp and significant rise in effective viscosity is observed (see \autoref{Fig:Viscosity_ratio}), leading to a more pronounced impact on the wall friction coefficient.

From a thermal perspective, the bulk temperature decreases along the axial direction. At any given axial location, the bulk temperature is lower for higher nanoparticle concentrations, as illustrated in \autoref{fig:Tb_B_10}. This behaviour occurs due to the rise in the thermal conductivity of the nanofluid(see in \autoref{Fig:Conductivity_ratio}). The reduction in bulk temperature along the axial direction indicates the development of the thermal boundary layer. Consequently, the thermal boundary layer thickness increases with increasing $\phi$ for both non-aggregation and aggregation cases. However, in the aggregation case, clustering of nanoparticles enlarges the effective surface area of the nanoparticles, leading to a steeper increase in thermal conductivity. As a result, heat from the wall is transferred to the fluid more rapidly than in the non-aggregation case, causing the bulk temperature to decrease more rapidly along the axial direction.

\begin{figure}[htbp]
    \centering

    \begin{subfigure}[b]{0.48\linewidth}
        \centering
        \includegraphics[width=\linewidth]{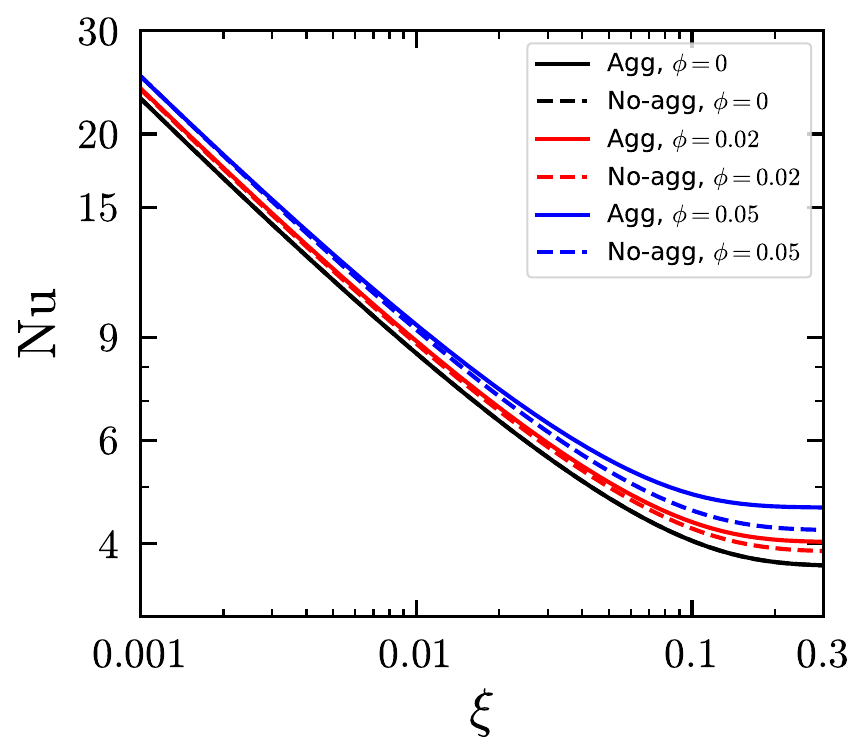}
        \caption{$B_n=0$}
    \end{subfigure}
      \hfill
     \begin{subfigure}[b]{0.48\linewidth}
        \centering
        \includegraphics[width=\linewidth]
  {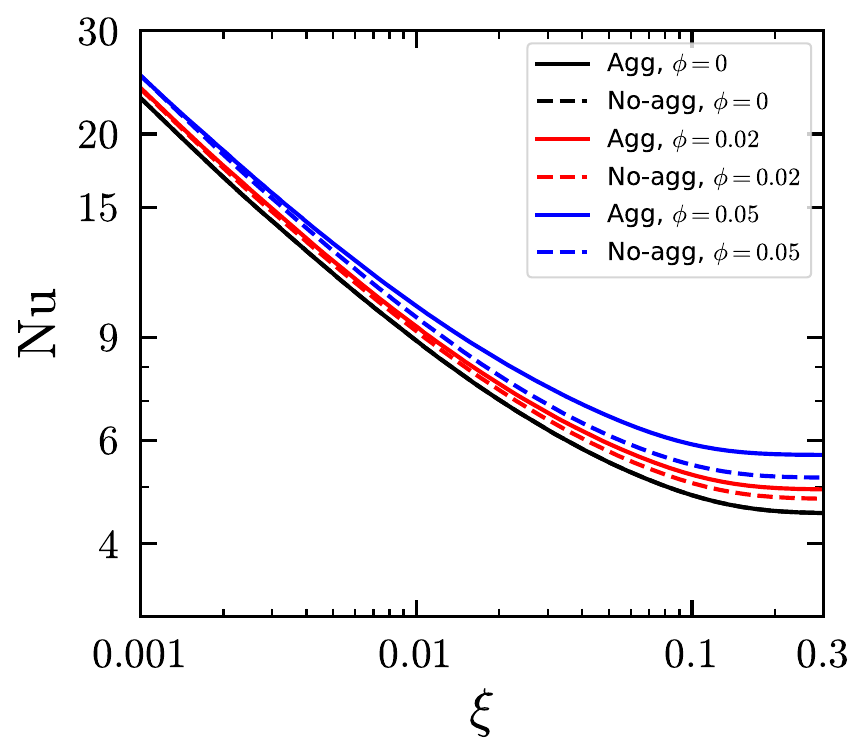}
        \caption{$B_n=10$}
    \end{subfigure}
    \vspace{0.3cm}
\begin{subfigure}[b]{0.48\linewidth}
        \centering
        \includegraphics[width=\linewidth]{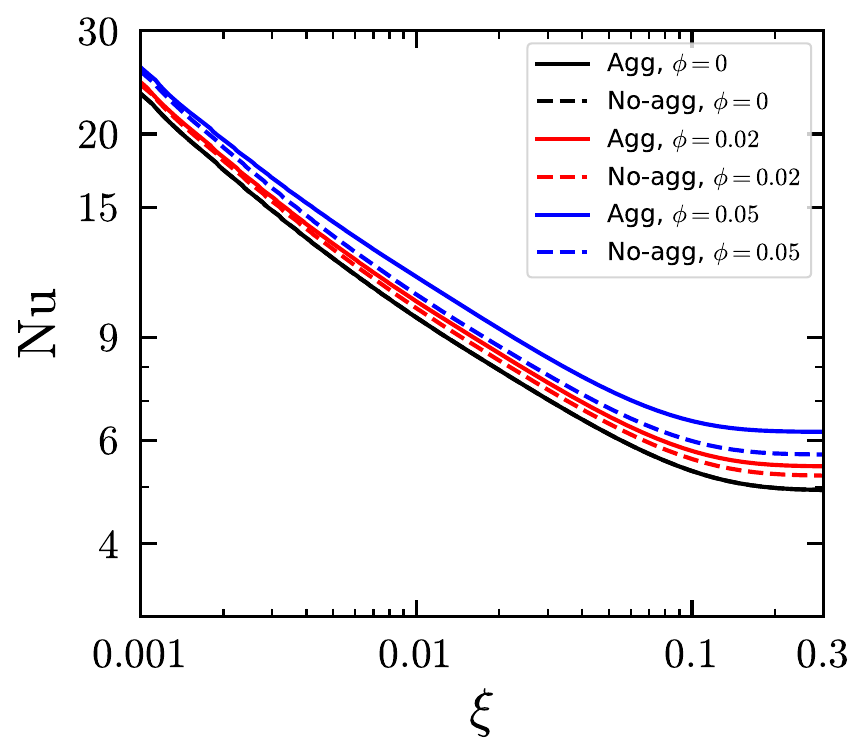}
        \caption{$B_n=30$}
    \end{subfigure}
    \hfill
     \caption{Variation of Nusselt Number along the axial direction for different values of Volume fraction considering both non-aggregation and aggregation models for $B_n = 0, 10, 30$.}
    \label{Fig:Nu_diff_B}
\end{figure}

The variation in bulk temperature will influence the Nusselt number. According to Eq.\eqref{eq:Non_dim_Nu}, as shown in \autoref{fig:Nu_B_10}, the local Nusselt number decreases rapidly along the axial direction in the inlet region, while towards the end of the developing region, it approaches a constant value. This pattern is consistent across all nanoparticle volume fractions. However, a slight increase in the Nusselt number is observed with higher nanoparticle concentrations. This enhancement in heat transfer can be attributed to improved thermal performance, especially in the case of nanoparticle aggregation, which leads to more significant heat transfer compared to the non-aggregation scenario, as quantified in \autoref{Table_Nusselt_number_phi_effect}. A similar trend is observed for the variation of Nusselt number for different Bingham numbers in \autoref{Fig:Nu_diff_B}.

The results include both aggregation and non-aggregation cases. \autoref{Table_Nusselt_number_phi_effect} summarises the numerical values of the friction coefficient at the wall and Nusselt number of a viscoplastic nanofluid at the cylinder inlet for different values of $\phi$ at different axial locations $\xi=$ 0.001, 0.01, and 0.3. The results clearly show that the influence of the nanoparticle volume fraction is more noticeable in the aggregation case than in the non-aggregation case.

\begin{table}[htbp]
\centering
\caption{Variation of skin friction coefficient and Nusselt number along the cylinder for different nanoparticle volume fractions $(\phi)$ and axial locations at $B_n = 10$ using both non-aggregation and aggregation models.}
\renewcommand{\arraystretch}{1.2}
\resizebox{\textwidth}{!}{
\begin{tabular}{|c|c|c|c|c|c|c|}
\hline
\multirow{2}{*}{\(\text{$B_n$}\)} & \multirow{2}{*}{\(\phi\)} & \multirow{2}{*}{\(\xi\)}
& \multicolumn{2}{c|}{Friction Coefficient ($C_f R_e$)} 
& \multicolumn{2}{c|}{Nusselt Number ($Nu$)} \\
\cline{4-7}
& & & Non-Aggregation & Aggregation & Non-Aggregation & Aggregation \\
\hline

\multirow{9}{*}{10} 

& \multirow{3}{*}{0.00} 
&0.001  & 96.4381 &  96.4381    &23.0539  &23.0539\\
&     & 0.01 &  70.0716 &70.0716   & 8.8832 &  8.8832\\
&     &  0.3  & 66.3229  & 66.3229 &4.5104   &  4.5104 \\
\cline{2-7}
& \multirow{3}{*}{0.01} 
& 0.001 & 98.9852   &  105.7829   &  23.4791   &  23.5364    \\
&     & 0.01  &  71.8781    & 77.8251    &   9.0520   &  9.1375   \\
&    & 0.3   &  68.0105   &  74.0493    & 4.6399  &  4.7265  \\
\cline{2-7}

& \multirow{3}{*}{0.02} 
& 0.001 &  101.6009   &  116.8588    &  23.9049  &  23.9947  \\
&    & 0.01  &  73.7496   &  87.1215   &  9.2227   &  9.3909    \\
&     & 0.3   &  69.7587   &  83.3950    &  4.7721   &  4.9501    \\
\cline{2-7}


& \multirow{3}{*}{0.05}
& 0.001 &  109.8938   & 167.0035   &   25.1864  &  25.1862    \\
&     & 0.01  &  79.7510  &  130.8731   &  9.7425   & 10.1795   \\
&   & 0.3   &  75.3971   &  127.6362   &  5.1852   &  5.6681   \\
\hline
\end{tabular}
}

\label{Table_Nusselt_number_phi_effect}
\end{table}

\subsection{\textbf{Effect of Bingham number on the flow characteristics of nanofluid:}}
\label{Sec:Effect_of_Bn}
In this subsection, we analyze the impact of the yield stress behavior of a viscoplastic nanofluid, considering three different Bingham numbers: 0, 10, and 30, with a nanoparticle volume fraction of 3\%.

\begin{figure}[htbp]
        \centering
        \includegraphics[width=1.0\linewidth]{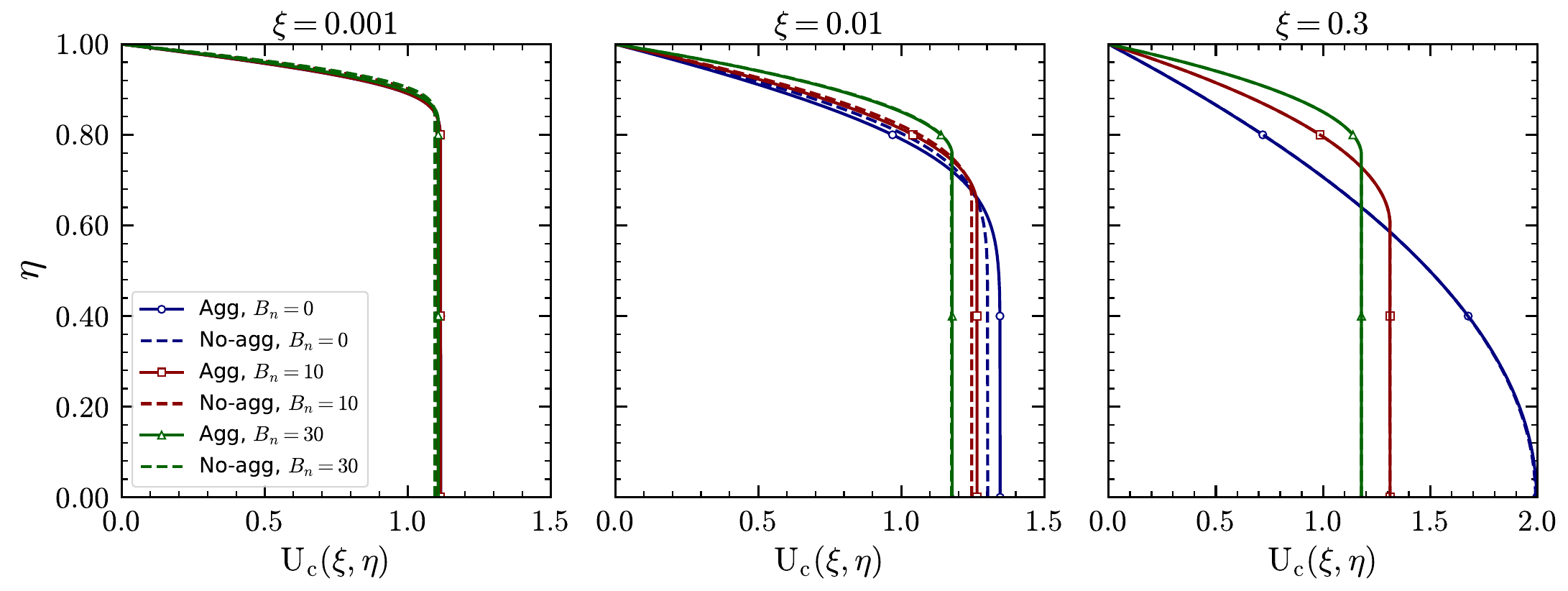}
        \caption{Development of axial velocity along the radial direction at axial locations 0.001, 0.01, and 0.3}
        \label{fig:Devel_U}
\end{figure}

\begin{figure}[htbp]
    \centering
    \includegraphics[width=0.65\linewidth]{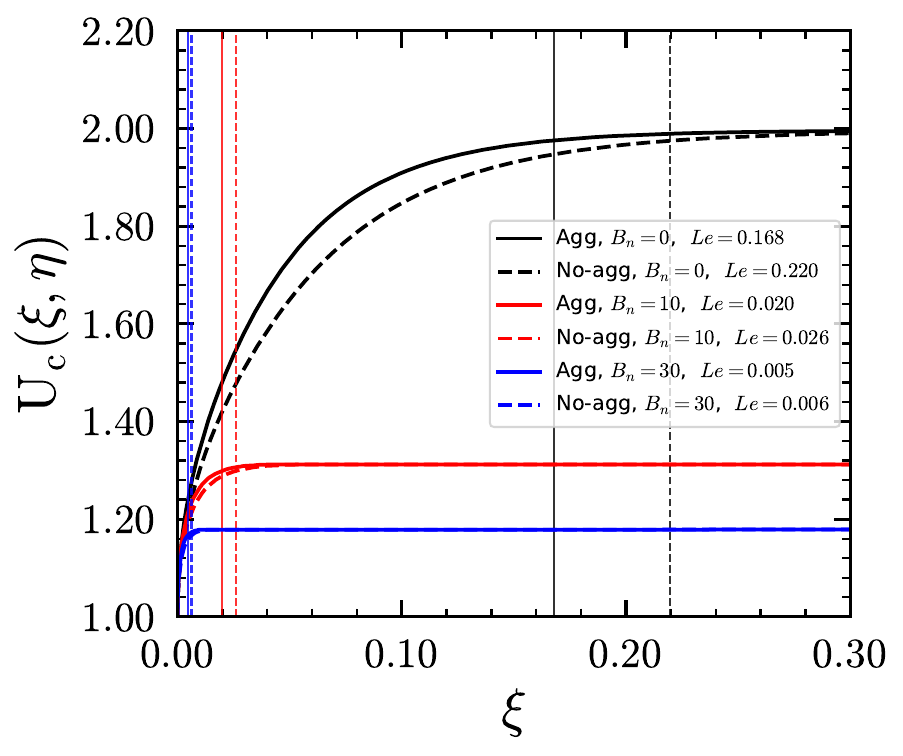}
    \caption{Variation of centerline axial velocity along the axial direction for various Bingham numbers with volume fraction $\phi=0.03$, comparing non-aggregation and aggregation models, with indication of the entry length ($Le$).}
    \label{fig:center_U_phi_0.03}
\end{figure}

The development of axial velocity in the cylinder for different Bingham numbers under both non-aggregated and aggregated nanoparticle conditions is illustrated in Figure \autoref{fig:Devel_U}. The non-dimensional axial plug flow velocity of the nanofluid at a nanoparticle volume fraction of \(\phi=0.03\) and Bingham numbers ranging from 0 to 30 exhibit a similar overall trend in both aggregation and non-aggregation scenarios. However, noticeable differences in velocity are observed in the inlet and developing regions between the aggregation and non-aggregation cases. Specifically, in the developing region, the plug flow or centerline velocity increases under aggregation conditions due to a rise in the nanofluid's apparent viscosity. This increase in viscosity accelerates the development of the boundary layer, allowing the velocity profile to reach its fully developed state over a shorter entry length (defined as the axial distance where 99\% of the fully developed velocity is attained). This effect is illustrated in \autoref{fig:center_U_phi_0.03} for a volume fraction of $\phi=0.03$. 

\begin{figure}[htbp]
    \centering
    \begin{subfigure}[b]{0.48\linewidth}
        \centering
        \includegraphics[width=\linewidth]{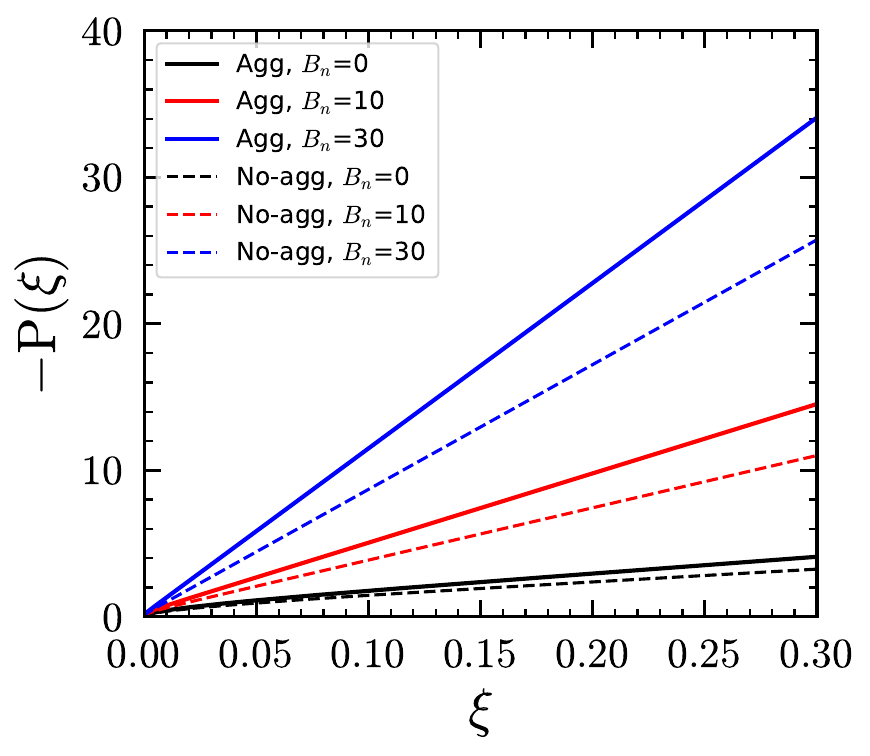}
        \caption{Pressure drop}
        \label{fig:pressure_phi_0_03}
    \end{subfigure}
      \hfill
        \begin{subfigure}[b]{0.48\linewidth}
        \centering
        \includegraphics[width=\linewidth]{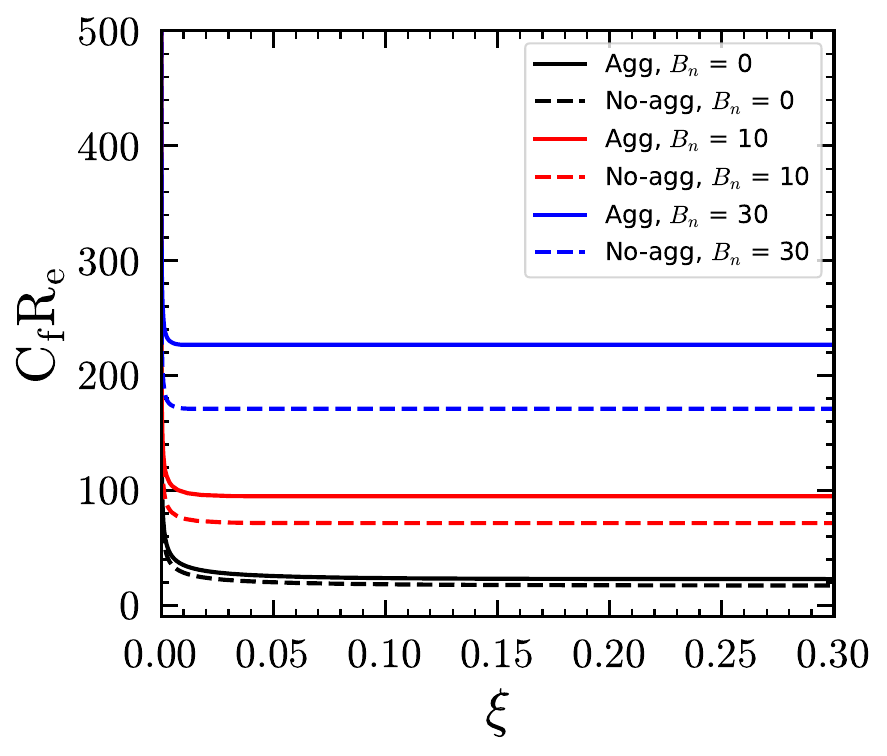}
        \caption{Friction coefficient variation}
        \label{fig:cf_phi_0_03}
    \end{subfigure}
    \vspace{0.3cm}

    \begin{subfigure}[b]{0.48\linewidth}
        \centering
        \includegraphics[width=\linewidth]{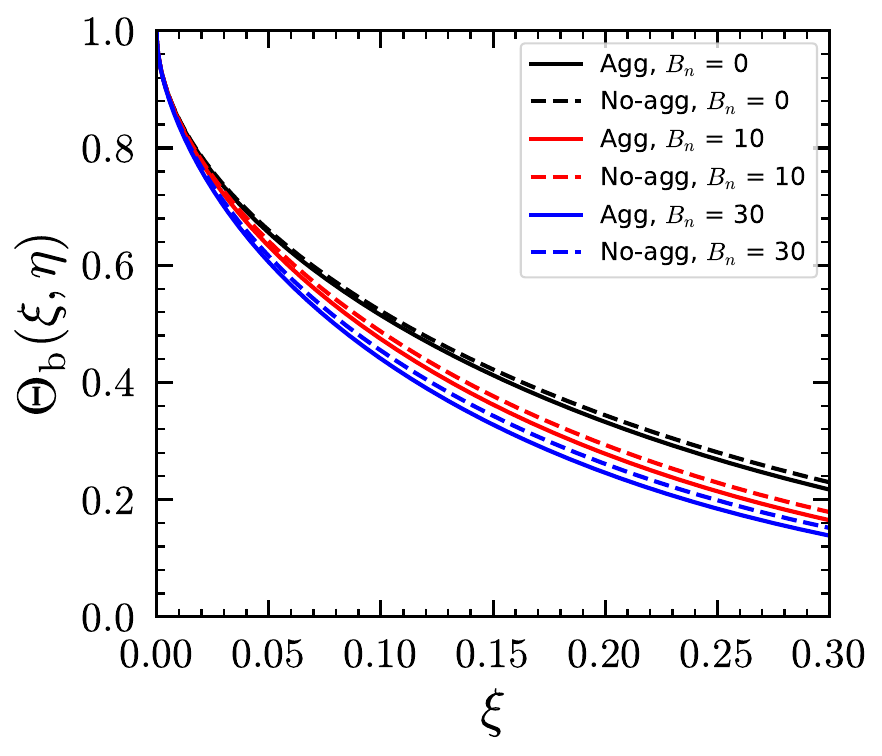}
        \caption{Bulk temperature}
       \label{fig:bulk_phi_0_03}
    \end{subfigure}
    \hfill
    \begin{subfigure}[b]{0.48\linewidth}
        \centering
        \includegraphics[width=\linewidth]{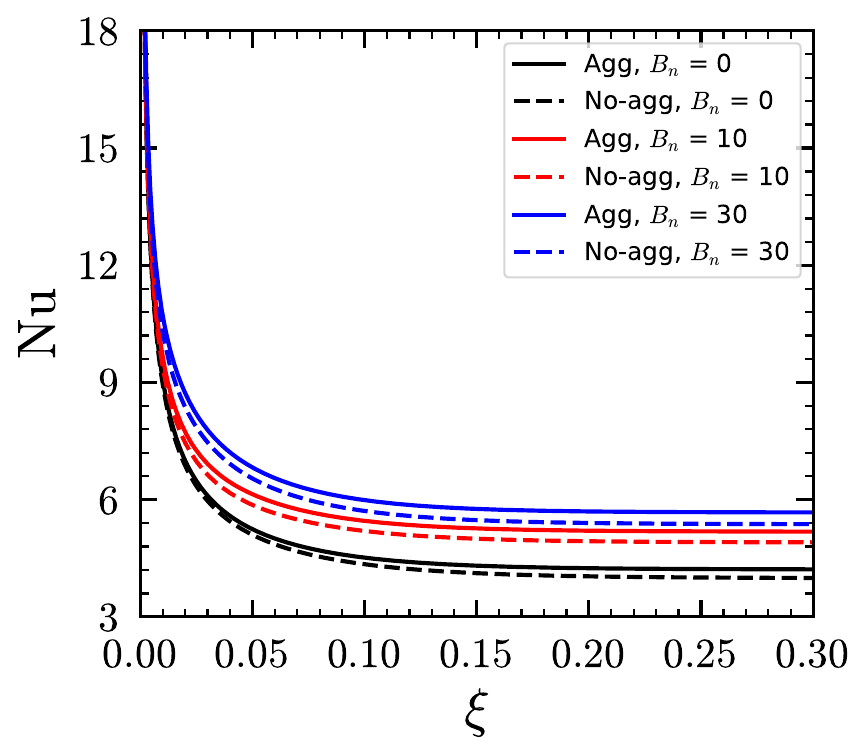}
        \caption{Nusselt number variation}
         \label{fig:Nu_phi_0_03}
    \end{subfigure}

    \caption{Variation of Pressure drop (\autoref{fig:pressure_phi_0_03}), Friction coefficient (\autoref{fig:cf_phi_0_03}), Bulk temperature (\autoref{fig:bulk_phi_0_03}) and Nusselt number (\autoref{fig:Nu_phi_0_03}) along the axial direction for different values of Bingham numbers considering both non-aggregation and aggregation models with volume fraction $\phi=0.03$.}
    \label{fig:combined_phi003}
\end{figure}

\autoref{fig:pressure_phi_0_03} presents the variation of pressure along the axial direction ($\xi$) for different Bingham numbers at a fixed nanoparticle volume fraction of $\phi=0.03$, considering both aggregation and non-aggregation cases. Consistent with earlier literature \cite{baioumy2021bingham} for the base fluid, a similar trend is observed for the nanofluid, where the pressure drop increases with increasing Bingham number due to the enhanced yield stress of the fluid. Furthermore, the pressure drop is higher in the aggregation case compared to the non-aggregation case due to the enlargement of the surface of the aggregated nanoparticles, which increases the flow resistance and consequently leads to a greater pressure drop.

\autoref{fig:cf_phi_0_03} illustrates the axial variation of the wall friction coefficient along the cylinder. From Eq. \eqref{eq:Non_dim_Cf} The friction coefficient, representing the wall shear stress, attains a peak magnitude in the inlet region due to the pronounced velocity gradient at the wall. As the flow develops downstream, the near-wall velocity gradient becomes progressively weaker, resulting in a corresponding reduction in the wall shear stress. Consequently, the friction coefficient gradually decreases along the axial direction until the flow reaches a fully developed state, where it becomes nearly constant. Fluids with higher yield stress exhibit higher apparent viscosity. At higher Bingham numbers, a sharper velocity gradient develops in the near-wall region, thereby intensifying the wall shear stress. Consequently, it increases friction coefficients at a given Reynolds number. In addition, nanoparticle aggregation further amplifies the frictional characteristics compared to the non-aggregated case. Specifically, the value of $C_f R_e$  for aggregated nanoparticles in a viscoplastic fluid is approximately 32.5\% greater than that of the non-aggregated case at 
 $\phi=0.03$.

The bulk temperature represents the thermal energy of the fluid within the cylinder. \autoref{fig:bulk_phi_0_03} shows the influence of yield stress (Bingham number) on the bulk temperature of the nanofluid along the axial direction. Fluids with higher Bingham numbers exhibit a faster loss of thermal energy due to stronger wall friction, which enhances heat transfer at the wall and leads to an earlier development of the thermal boundary layer. Moreover, in the aggregation case, the thermal boundary layer develops more rapidly than in the non-aggregation case. This variation of bulk temperature directly impacts the Nusselt number, which measures heat transfer.

\begin{figure}[htbp]
    \centering

    \begin{subfigure}[b]{0.48\linewidth}
        \centering
          \includegraphics[width=\linewidth]{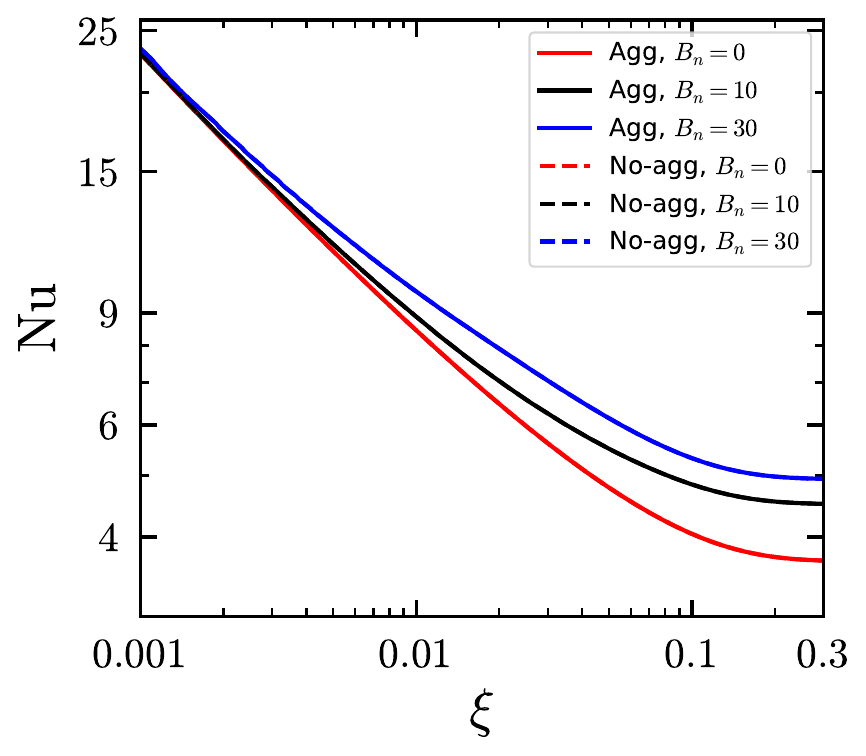}
        \caption{$\phi=0.0$}
        \label{main_plor_0}
    \end{subfigure}
      \hfill
     \begin{subfigure}[b]{0.48\linewidth}
        \centering
        \includegraphics[width=\linewidth] {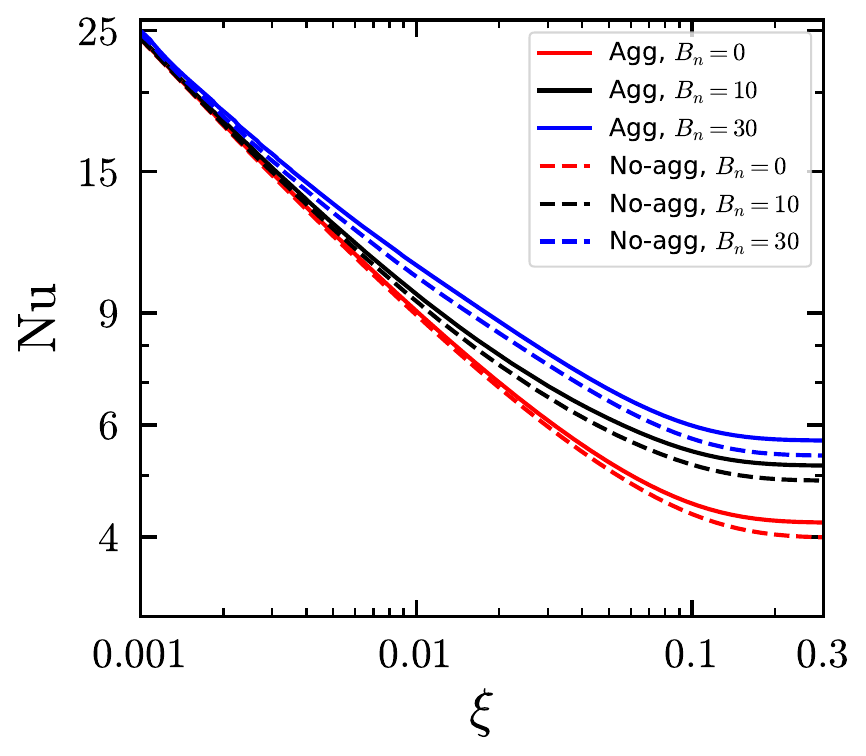}
       \caption{$\phi=0.03$}
        \label{main_plor_0_03}
    \end{subfigure}
    \vspace{0.3cm}

      \begin{subfigure}[b]{0.48\linewidth}
        \centering
        \includegraphics[width=\linewidth]{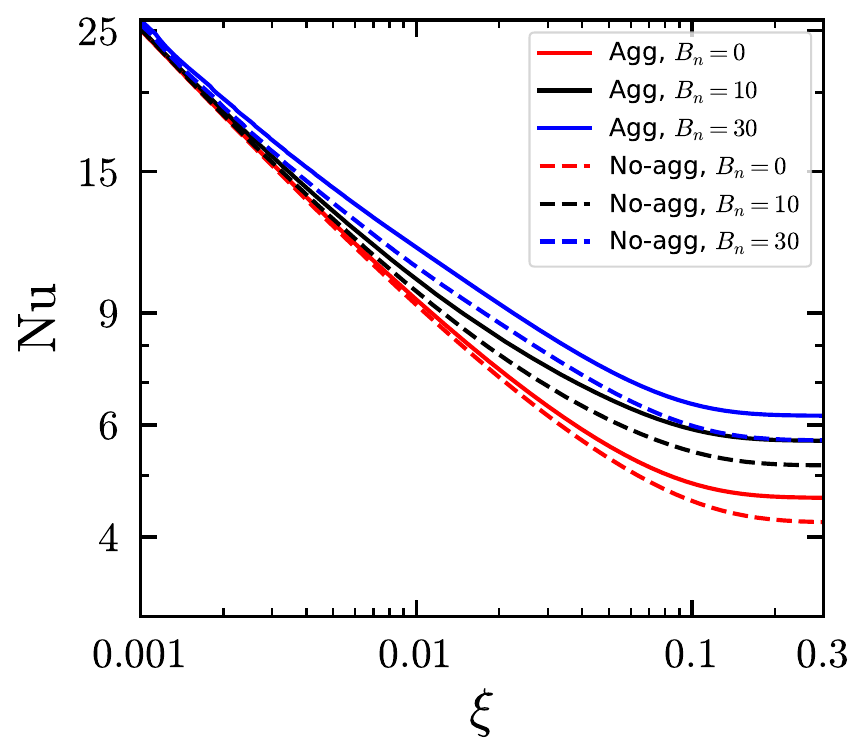}
          \caption{$\phi=0.05$}
        \label{main_plor_0_05}
    \end{subfigure}
    \hfill
    \caption{Variation of Nusselt Number along the axial direction for different values of Bingham numbers considering both non-aggregation and aggregation models for $\phi = 0, 0.03, 0.05$.}
    \label{Fig:Nu_diff_phi}
\end{figure}

\autoref{fig:Nu_phi_0_03} shows the variation of the Nusselt number along the axial direction. The Nusselt number rapidly decreases along the flow direction, with the influence of the Bingham number becoming more significant in the developing region than near the inlet. In the inlet region, the variation in Nusselt number with respect to the Bingham number is minimal because there is minimal change in the temperature wall gradient; however, in the developing region, fluids with higher yield stress exhibit larger Nusselt numbers due to the increased temperature gradient at the wall and faster drops in bulk temperature from Eq.\eqref{eq:Non_dim_Nu}. The Nusselt number is higher for nanofluids containing aggregated nanoparticles due to higher effective thermal conductivity. At a volume fraction of $\phi=0.03$, the rate of increase in the Nusselt number for the aggregation case, compared to the non-aggregation case, is approximately $5.6\%$ at the beginning of the developed region for all viscoplastic nanofluids. Similarly, we demonstrated the variation of the Nusselt number for viscoplastic nanofluid with different volume fractions along the axial direction in \autoref{Fig:Nu_diff_phi}, and we have quantified the results, which can be found in \autoref{Table_Nusselt_number_Bn_effect}.

\autoref{Table_Nusselt_number_Bn_effect}  presents the computational results for different Bingham numbers at a fixed value of $\phi=0.03$. These results reflect the characteristic behaviour of viscoplastic fluids, in which higher Bingham numbers correspond to larger values of both the friction coefficient and the Nusselt number. The improvement is more pronounced in the aggregated case than in the non-aggregated case.

\begin{table}[htbp]
\centering
\caption{Variation of skin friction coefficient and Nusselt number along the cylinder for different Bingham number $(B_n)$ and axial locations at $\phi=0.03$ using both non-aggregation and aggregation models.}
\renewcommand{\arraystretch}{1.25}
\begin{tabular}{|c|c|c| c|c| c|c|}
\hline
\multirow{2}{*}{$\phi$} & \multirow{2}{*}{$B_n$} & \multirow{2}{*}{$\xi$} 
& \multicolumn{2}{c}{Friction Coefficient ($C_f R_e$)} 
& \multicolumn{2}{c|}{Nusselt Number ($Nu$)} \\ 
\cline{4-7}
& & & Non-Aggregation & Aggregation & Non-Aggregation & Aggregation \\
\hline

\multirow{9}{*}{0.03}

& \multirow{3}{*}{0}
& 0.001 & 60.7148 & 72.3804 & 24.2872& 24.3695 \\
& & 0.01 &28.4462 & 35.0397 & 8.9413 & 9.0696 \\
& & 0.3  & 17.2729 & 22.8546& 3.9952 & 4.2155\\
\cline{2-7}

& \multirow{3}{*}{10}
& 0.001 & 104.2886& 130.1900& 24.3312 &24.4255\\
& & 0.01 & 75.6985 & 98.5044 & 9.3957 &  9.6480 \\
& & 0.3  & 71.5706 & 94.8778 & 4.9070 & 5.1812 \\
\cline{2-7}
& \multirow{3}{*}{30}
& 0.001 & 193.8199& 249.9290& 24.7527 & 25.0237\\
& & 0.01 & 171.3600 & 226.7733 & 10.2756 & 10.7049 \\
& & 0.3  & 171.0471 & 226.7494& 5.3727&  5.6731 \\
\hline
\end{tabular}
\label{Table_Nusselt_number_Bn_effect}
\end{table}

\subsection{\textbf{Performance evaluation criteria(PEC):}}
In summary, the heat transfer enhancement of the viscoplastic nanofluid with increasing nanoparticle volume fraction is primarily attributed to the higher thermal conductivity of the nanoparticles. However, at higher volume fractions, nanoparticles tend to cluster together, leading to aggregation—a phenomenon where nonmaterial group collectively rather than remaining uniformly dispersed. In the absence of aggregation, nanoparticles are assumed to be homogeneously dispersed. 

The comparative results reveal that considering nanoparticle aggregation significantly enhances the predicted heat transfer performance compared to the non-aggregation scenario, owing to the greater increase in both viscosity and thermal conductivity associated with aggregation effects.

In this regard, we use the performance evaluation criteria (PEC) to assess the efficiency of adding nanoparticles to a base fluid. It quantifies the overall impact of nanoparticle addition by considering both the friction factor and the Nusselt number. Inclusion of nanoparticles generally increases the fluid’s viscosity, leading to a higher friction factor and pressure drop, while simultaneously enhancing the thermal conductivity, which improves heat transfer and increases the Nusselt number. A PEC value greater than one indicates an effective and beneficial use of nanoparticles. The PEC is expressed as~\cite{najafabadi2024entry}:

\begin{equation}
     PEC=\frac{Nu_{nf}}{Nu_{bf}} \left(\frac{{C_f}_{nf}}{{C_f}_{bf}}\right)^{-\frac{1}{3}}
     \label{Eq:PEC}
\end{equation}

\begin{figure}[H]
    \centering
    \includegraphics[width=0.6\linewidth]{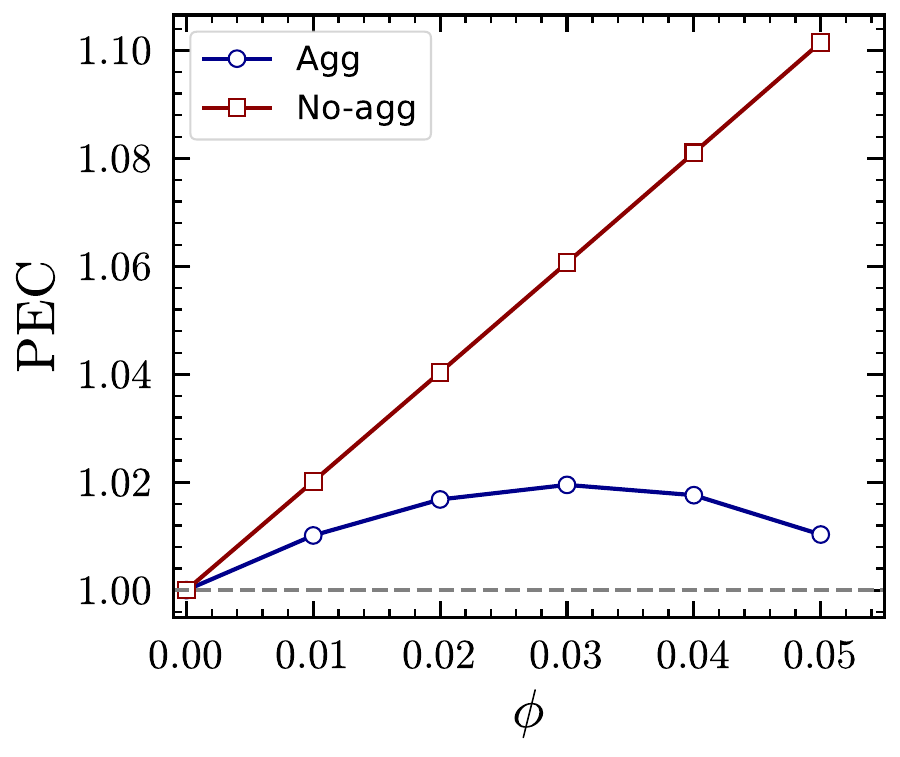}
    \caption{Performance evaluation criteria for change in volume fraction at Bingham number, $B_n = 10$ and $Pr=1$ for both non-aggregation and aggregation models}
    \label{fig:PEC_0_05}
\end{figure}

We have evaluated the Performance Evaluation Criteria (PEC) for both aggregated and non-aggregated nanoparticles, as shown in \autoref{fig:PEC_0_05}, considering a viscoplastic nanofluid with a Bingham number of 10 and nanoparticle volume fractions up to 5\%. The results reveal that the PEC of viscoplastic nanofluids is efficient and advantageous up to a 5\% volume fraction.

\begin{figure}[H]
    \centering
\includegraphics[width=0.6\linewidth]{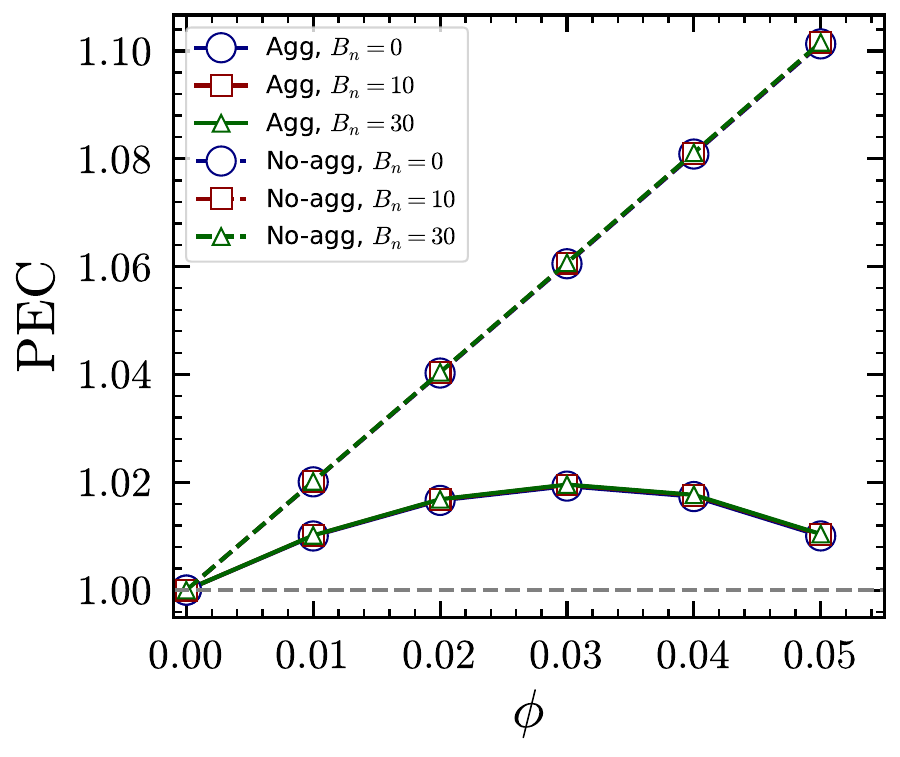}
    \caption{Performance evaluation criteria for change in volume fraction at Bingham number, $B_n = 0,10,30$ and $Pr=1$ for both non-aggregation and aggregation models}
    \label{fig:PEC_all_Bn_0_05}
\end{figure}

In the case of non-aggregated nanoparticles, the PEC increases steadily with increasing volume fraction. However, for aggregated nanoparticles, the PEC remains greater than one for all volume fractions — it initially increases with volume fraction up to 3\% and then begins to decrease beyond that point. This behaviour can be attributed to the fact that PEC depends on the ratio of heat transfer enhancement to the friction coefficient. As the nanoparticle volume fraction increases, both density and viscosity rise, leading to a higher friction factor. Consequently, beyond 3\%, the increase in friction dominates, reducing the PEC. Therefore, the optimal performance for aggregated nanoparticles in viscoplastic nanofluids occurs at a 3\% volume fraction, where the system exhibits peak efficiency.
Similarly, we have also analyzed the PEC values for different Bingham numbers, as shown in \autoref{fig:PEC_all_Bn_0_05}. For the aggregation case, the peak PEC value occurs at a volume fraction of 3\%. In contrast, for the non-aggregation case, the PEC value increases as the volume fraction of nanoparticles increases, as shown in \autoref{Table:PEC_all_Bn}.

\begin{table}[htbp]
\centering
\caption{Performance evaluation criteria (PEC) for different $\phi$ and Bingham numbers}
\label{Table:PEC_all_Bn}
\begin{tabular}{|c|ccc|ccc|}
\hline
\multirow{2}{*}{$\phi$} & \multicolumn{3}{c|}{Aggregation} & \multicolumn{3}{c|}{Non-Aggregation} \\ \cline{2-7}
 & $B_n=0$ & $B_n=10$ & $B_n=30$ & $B_n=0$ & $B_n=10$ & $B_n=30$ \\
\hline
0.00 & 1.0000 & 1.0000 & 1.0000 & 1.0000 & 1.0000 & 1.0000 \\
0.01 & 1.0100 & 1.0101 & 1.0102 & 1.0201 & 1.0201 & 1.0202 \\
0.02 & 1.0166 & 1.0168 & 1.0169 & 1.0403 & 1.0404 & 1.0404 \\
0.03 & 1.0193 & 1.0195 & 1.0196 & 1.0605 & 1.0607 & 1.0607 \\
0.04 & 1.0173 & 1.0176 & 1.0177 & 1.0809 & 1.0810 & 1.0811 \\
0.05 & 1.0100 & 1.0103 & 1.0104 & 1.1013 & 1.1015 & 1.1016 \\
\hline
\end{tabular}
\end{table}

\section{Conclusion}
This study numerically investigated the steady, laminar viscoplastic nanofluid flow in the entrance region of a circular cylinder with uniform wall temperature, considering the effects of both nanoparticle non-aggregation and aggregation. The Brinkman and Maxwell models are used for non-aggregation, while the Krieger–Dougherty and Maxwell–Bruggeman models are applied for aggregation. The key findings are summarized below.

\begin{itemize}
    \item Nanoparticle addition does not affect the velocity profile in the fully developed region, but alters it in the developing region. Aggregated nanoparticles yield higher velocities than non-aggregated ones.
    \item The pressure drop grows proportionally with the Bingham number, and the presence of nanoparticles further amplifies it at each Bingham number. For nanofluids with volume fractions of 0.03 and 0.05 at $B_n=10$, the pressure drop rises by 7.93\% and 13.71\% in the non-aggregation case, and by 42.35\% and 90.82\% in the aggregation case, respectively, compared to the base fluid.
    
    The pressure drop increase raises the friction coefficient, which grows with nanoparticle volume fraction and is significantly higher under aggregation.

    \item Near the inlet, the Nusselt number is weakly influenced by the Bingham number but grows with the inclusion of nanoparticles, declines along the flow direction, whereas in the developing region, it depends on both the Bingham number and volume fraction.

    At $B_n=10$, it increases by 8.79\% and 14.96\% (non-aggregation) and 14.87\% and 25.67\% (aggregation) for volume fractions of 0.03 and 0.05, respectively, compared to the base fluid.

    \item Performance evaluation criterion (PEC) was examined for various Bingham numbers and nanoparticle volume fractions. The results show that it remains above 1 within the volume fraction range of 0–5\% for both aggregation and non-aggregation conditions. Whereas, in the non-aggregation case, it soars linearly with volume fraction, whereas in the aggregation case, it diminishes at higher volume fractions, reaching a maximum at 3\% volume fraction.

\end{itemize}

\section{Limitations and Future Scope}
The study conducted in this work is about heat transfer and flow in the entrance region. The parameters considered in this study are mainly the Bingham number (representing the yield stress of the base fluid) and the Volume fraction (representing the ratio of the volume of nanoparticles in the base fluid), and its effects on the flow characteristics (such as pressure drop and friction factor) and thermal characteristics (such as bulk temperature and Nusselt number). Furthermore, the nanofluids include the effects of non-aggregation and aggregation of nanoparticles using single-phase models for viscosity and thermal conductivity. The current study has several limitations that should be addressed in future research.
\begin{enumerate}
    \item The study does not focus on the effects of the Reynolds number and Prandtl numbers. Hence, to gain a better understanding of the flow and thermal characteristics, the effects of the Reynolds number and Prandtl number may be included. PEC may further be analyzed including the effects of Reynolds number and Prandtl number.
    \item The study is based on single-phase nanofluid models for both non-aggregation and aggregation cases. These models have a limitation of the range of volume fractions, and hence the study conducted using single-phase models is based on effective parameters (such as effective viscosity, density ratio, specific heat, and thermal conductivity). These parameters indicate the results in an \textit{effective}/macroscopic sense. Hence, simulations using two-phase flow involving coupled equations pertaining to the flow and particle motion may provide better information in the sheared/yielded/boundary layer regions, especially when aggregation of particles is involved.
    \item The study can be extended to include effects such as magnetic field, Joule heating, Brownian motion, thermophoresis, etc using appropriate models (such as Carreau model, Sisko model for base fluids \cite{khan2026computational,khan2026distributed} and disperse hybrid nanoparticles \cite{irfan2025thermal}), boundary conditions, and computational methods (such as the Keller Box\cite{muhammad2025significance} method).

\end{enumerate}

\appendix
\renewcommand{\thesection}{Appendix \Alph{section}}
\section{ Grid Convergence Index and Validation Results}
\subsection{Grid Convergence Index (GCI)}
\label{Sec:GCI}

In this subsection, we compute the grid convergence indices for different choices of grid sizes. The GCI is computed based on Richardson extrapolation \cite{celik2008procedure} as:

\begin{equation}
GCI_{21} = F_s 
\frac{\left| \dfrac{f_1 - f_2}{f_1} \right|}
{r_{21}^{\,p} - 1} \times 100
\end{equation}

\noindent
where $f_1$, $f_2$, and $f_3$ denote the solutions obtained on the fine, medium, and coarse grids, respectively.  
The refinement ratio is defined as

\begin{equation}
r_{21} = \frac{h_2}{h_1},
\end{equation}

\noindent
with $h$ representing the characteristic grid spacing.  
The observed order of accuracy $p$ is calculated as

\[
p =\frac{1}{\ln(r_{21})} \left|
\ln\left| \frac{f_3 - f_2}{f_2 - f_1} \right|
+ \ln\left( \frac{r_{32}^p - s}{r_{21}^p - s} \right)\right|
\]
\[
\text{where } s = \text{sign}\left( \frac{f_3 - f_2}{f_2 - f_1} \right)
\]
\noindent
Here, $F_s$ is the safety factor (taken as 1.25 for three-grid studies).  
The GCI quantifies the percentage discretisation error of the fine-grid solution. 

We utilised different grid resolutions: $L_1$: ($8471 \times 256$), $L_2$: ($11857 \times 358$), and $L_3$: ($16601 \times 501$), designated as the coarse grid, medium grid, and fine grid, respectively.

\begin{table}[htbp]
    \centering
    \caption{GCI for Centerline Velocity $U(\xi,\eta)$ at $\phi = 0.04$,$B_n=10$, C-Ratio $=\displaystyle{\frac{GCI_{32}}{GCI_{21}r^{p}}}$}
    \label{tab:gci_study_U_B_10}
    \small
    \begin{tabular}{ccccccccc}
        \toprule
        Target $x$ & $U_{fine}$ & $U_{med}$ & $U_{coarse}$ & $r_{21}$ & $p$ & GCI$_{21}$ (\%) & GCI$_{32}$ (\%) & C-Ratio \\
        \midrule
        0.001      & 1.120660  &1.119153  & 1.116789 & 1.4 & 1.340 & 0.2950    &0.4637    & 1.0013 \\ 
        0.15       & 1.312117  & 1.311189  & 1.308767  & 1.4  & 2.852 & 0.0548    & 0.1433    & 1.0007 \\ 
        0.3       & 1.312117 & 1.311189  &1.308767  & 1.4 &2.852  & 0.0548    &0.1433   & 1.0007\\ 
        \bottomrule

         \\
    \end{tabular}
\end{table}

\begin{table}[htbp]
    \centering
    \caption{GCI for local Nusselt number $Nu$ at $\phi = 0.04$, $B_n=10$, C-Ratio $=\displaystyle{\frac{GCI_{32}}{GCI_{21}r^{p}}}$}
    \label{tab:gci_study_Nu_B_10}
    \small
    \begin{tabular}{ccccccccc}
        \toprule
        Target $x$ & $Nu_{fine}$ & $Nu_{med}$ & $Nu_{coarse}$ & $r_{21}$ & $p$ & GCI$_{21}$ (\%) & GCI$_{32}$ (\%) & C-Ratio \\
        \midrule
        0.001      & 24.824078 & 24.876383 & 24.950550 & 1.4  & 1.039 &0.6286    &0.8901    & 0.9979 \\ 
        0.15       &5.492690 &5.496535  & 5.505635 & 1.4  & 2.562  & 0.0639    &0.1513    & 0.9993 \\ 
        0.3        & 5.420462  & 5.424383  & 5.433478 & 1.4  &2.502  &0.0684   & 0.1587    &0.9993 \\

        \bottomrule
\end{tabular}
\end{table}

According to the GCI analysis for velocity and Nusselt, from above \autoref{tab:gci_study_U_B_10} and \autoref{tab:gci_study_Nu_B_10}, the fine grid offers grid-independent solutions with discretisation errors below 1\% for our choice of $B_n=10, \phi=0.04$. Consequently, the fine grid has been chosen for further computations to strike an optimal balance between computational efficiency and numerical accuracy.

\subsection{Validation of results:}
\label{Sec_Validation}

The following tables provide information on the deviation between our results and earlier literature in the case of the friction factor at the wall and the Nusselt number.

\begin{table}[h]
\centering
\caption{Comparison of $C_f Re$ between the present study and Baioumy \textit{et al} \cite{baioumy2021bingham} at $B_n = 9.86$.}
\label{tab:validation_B04}
\begin{tabular}{cccc}
\toprule
$\xi$ &  Baioumy \textit{et. al} &Present Study  & Deviation (\%) \\
\midrule
0.005  &  74.301   &  73.797  &  0.678\\
0.009  &  69.747   &  70.148  &  0.575 \\
0.012  &  67.350   &  68.208  &  1.274\\
0.112  &  65.792   &  65.666  &  0.193 \\
0.220  &  65.792   &  65.666  &  0.193\\
0.294  &  65.792   &  65.666  &  0.193\\
\bottomrule
\end{tabular}
\end{table}

\begin{table}[H]
    \centering
\caption{Comparison of the local Nusselt number with the results of Benkhedda \textit{et al.} \cite{benkhedda2020convective} for $\phi = 0.00$.}
\begin{tabular}{cccc}
\toprule
$\xi$ & Benkhedda \textit{et al.} & Present study & Deviation (\%) \\
\midrule
0.0770 & 8.7500 & 9.0470 & 3.3947 \\
0.0955 & 8.2143 & 8.3749 & 1.9558 \\
0.1141 & 7.8571 & 7.8756 & 0.2347 \\
0.1369 & 7.3214 & 7.4087 & 1.1917 \\
0.1569 & 6.9643 & 7.0889 & 1.7897 \\
0.1811 & 6.6071 & 6.7766 & 2.5651 \\
0.2011 & 6.6071 & 6.5648 & 0.6403 \\
0.2253 & 6.2500 & 6.3484 & 1.5740 \\
0.2481 & 6.0714 & 6.1759 & 1.7209 \\
0.2667 & 6.0714 & 6.0534 & 0.2973 \\
0.2923 & 5.8929 & 5.9046 & 0.1989 \\
0.3123 & 5.8929 & 5.8026 & 1.5310 \\
\bottomrule
\end{tabular}
\label{tab:Nu_val_0.0}
\end{table}

\begin{table}[H]
 \centering
\caption{Comparison of the local Nusselt number with the results of Benkhedda \textit{et al.} \cite{benkhedda2020convective} for $\phi = 0.04$.}
\begin{tabular}{cccc}
\toprule
$\xi$ & Benkhedda \textit{et al.} & Present study & Deviation (\%) \\
\midrule
0.0699 & 9.6429 & 9.9922 & 3.6232 \\
0.0884 & 8.9286 & 9.1798 & 2.8138 \\
0.1112 & 8.2143 & 8.4766 & 3.1939 \\
0.1312 & 7.8571 & 8.0214 & 2.0909 \\
0.1512 & 7.5000 & 7.6623 & 2.1639 \\
0.1711 & 7.3214 & 7.3704 & 0.6691 \\
0.1911 & 7.1429 & 7.1277 & 0.2116 \\
0.2153 & 6.7857 & 6.8822 & 1.4225 \\
0.2367 & 6.6071 & 6.6997 & 1.4009 \\
0.2581 & 6.4286 & 6.5420 & 1.7639 \\
0.2852 & 6.2500 & 6.3702 & 1.9230 \\
0.3080 & 6.2500 & 6.2450 & 0.0807 \\
\bottomrule
\end{tabular}
\label{tab:Nu_val_0.04}
\end{table}
\unappendix

\section*{Acknowledgements}
The authors sincerely thank the anonymous reviewers for their comments and suggestions that have improved the quality of the discussions in the article. Deepa Madivalar acknowledges the use of PARAM UTKARSH HPC facility of the National Supercomputing Mission project and thanks the center for Cyber Physical Systems (CCPS), National Institute of Technology Karnataka, Surathkal for the financial support for the same. Deepa Madivalar also thanks National Institute of Technology Karnataka, Surathkal for providing the fellowship to pursue her PhD.

\bibliographystyle{elsarticle-num} 
\bibliography{cas-refs}

@article{saidur2011review,
  title={A review on applications and challenges of nanofluids},
  author={Saidur, Rahman and Leong, KY and Mohammed, Hussein A},
  journal={Renewable and sustainable energy reviews},
  volume={15},
  number={3},
  pages={1646--1668},
  year={2011},
  publisher={Elsevier}
}

@article{ma2023heat,
  title={Heat transfer enhancement of nanofluid flow at the entry region of microtubes},
  author={Ma, Hao and He, Boshu and Su, Liangbin and He, Di},
  journal={International Journal of Thermal Sciences},
  volume={184},
  pages={107944},
  year={2023},
  publisher={Elsevier}
}

@article{bhavya2025computational,
  title={Computational analysis of strontium stannate nanoparticles flow and non-Fourier-Fick’s heat flux over an expanding cylinder: A nanoparticles aggregation study},
  author={Bhavya, D and Vasanth, KR and Gowtham, HJ and Kumar, K Ganesh},
  journal={Journal of Molecular Liquids},
  volume={427},
  pages={127460},
  year={2025},
  publisher={Elsevier}
}

@article{shen2024entropy,
  title={Entropy optimization and heat transfer in thin film flow of electromagnetic micropolar nanofluid using Maxwell--Bruggeman and Krieger--Dougherty models},
  author={Shen, Shuifa and Rehman, Sohail and Shah, Syed Omar and Albouchi, Fethi and Rauf, Somiya},
  journal={Alexandria Engineering Journal},
  volume={106},
  pages={71--86},
  year={2024}
}

@article{baioumy2021bingham,
  title={Bingham fluid flow in the entrance region of a pipe},
  author={Baioumy, Basma and Chebbi, Rachid and Abdel Jabbar, Nabil},
  journal={Journal of Fluids Engineering},
  volume={143},
  number={2},
  pages={024503},
  year={2021},
  publisher={American Society of Mechanical Engineers}
}

@article{sadeghinezhad2016comprehensive,
  title={A comprehensive review on graphene nanofluids: Recent research, development and applications},
  author={Sadeghinezhad, Emad and Mehrali, Mohammad and Saidur, R and Mehrali, Mehdi and Latibari, Sara Tahan and Akhiani, Amir Reza and Metselaar, Hendrik Simon Cornelis},
  journal={Energy Conversion and Management},
  volume={111},
  pages={466--487},
  year={2016},
  publisher={Elsevier}
}

@article{choi1995enhancing,
    author = {Choi, S.U.S. and Eastman, Jeffrey},
    year = {1995},
    title = {Enhancing thermal conductivity of fluids with nanoparticles},
    volume = {66},
    journal = {Proceedings of the ASME International Mechanical Engineering Congress and Exposition}
}

@article{najafabadi2024entry,
  title={Entry length correlations for alumina-water nanofluid in laminar pipe flow},
  author={Najafabadi, Mohsen Khalili and Hricz{\'o}, Kriszti{\'a}n and Bogn{\'a}r, Gabriella},
  journal={International Journal of Thermal Sciences},
  volume={197},
  pages={108808},
  year={2024},
  publisher={Elsevier}
}

@article{teamah2018influence,
  title={Influence of nano-particles addition on hydrodynamics and heat transfer in laminar flow entrance region inside tube},
  author={Teamah, Ahmed M and Hassab, Mohamed A and El-Maghlany, Wael M},
  journal={Alexandria engineering journal},
  volume={57},
  number={4},
  pages={4091--4102},
  year={2018},
  publisher={Elsevier}
}

@article{dai2024mechanism,
  title={Mechanism of enhanced thermal conductivity of hybrid nanofluids by adjusting mixing ratio of nanoparticles},
  author={Dai, Jinghui and Zhai, Yuling and Li, Zhouhang and Wang, Hua},
  journal={Journal of Molecular Liquids},
  volume={400},
  pages={124518},
  year={2024},
  publisher={Elsevier}
}

@article{ali2024effect,
  title={Effect of Al2O3/H2O Nanofluid on the Flow and Forced Convection Heat Transfer Enhancement in a Pipe Using Commercial CFD Code},
  author={Ali, Sarmad A and Hameed, Mohanad R and Kadhim, Hanan K},
  journal={Iraqi Journal of Industrial Research},
  volume={11},
  number={3},
  pages={11--24},
  year={2024}
}

@article{maiga2005heat,
  title={Heat transfer enhancement by using nanofluids in forced convection flows},
  author={Maiga, Sidi El Becaye and Palm, Samy Joseph and Nguyen, Cong Tam and Roy, Gilles and Galanis, Nicolas},
  journal={International journal of heat and fluid flow},
  volume={26},
  number={4},
  pages={530--546},
  year={2005},
  publisher={Elsevier}
}

@article{labib2013numerical,
  title={Numerical investigation on effect of base fluids and hybrid nanofluid in forced convective heat transfer},
  author={Labib, M Nuim and Nine, Md J and Afrianto, Handry and Chung, Hanshik and Jeong, Hyomin},
  journal={International Journal of Thermal Sciences},
  volume={71},
  pages={163--171},
  year={2013},
  publisher={Elsevier}
}

@article{nadiminti2020heat,
  title={Heat and mass transfer effects of casson fluid in the entrance of concentric annuli with moviment of walls},
  author={Nadiminti, SR and Kandasamy, A},
  journal={Advances in Mathematics: Scientific Journal},
  volume={9},
  number={9},
  pages={6435--6446},
  year={2020}
}

@article{venthan2019analysis,
  title={Analysis of entrance region flow of Bingham nanofluid in concentric annuli with rotating inner cylinder},
  author={Venthan, Selvam Mullai and Amalraj, Isaac Jayakaran and Kumar, Ponnusamy Senthil},
  journal={Micro \& Nano Letters},
  volume={14},
  number={13},
  pages={1361--1365},
  year={2019},
  publisher={Wiley Online Library}
}

@article{hazeri2021three,
  title={Three-dimensional analysis of forced convection of Newtonian and non-Newtonian nanofluids through a horizontal pipe using single-and two-phase models},
  author={Hazeri-Mahmel, Naser and Shekari, Younes and Tayebi, Ali},
  journal={International Communications in Heat and Mass Transfer},
  volume={121},
  pages={105119},
  year={2021},
  publisher={Elsevier}
}

@article{akbari2017effect,
  title={The effect of velocity and dimension of solid nanoparticles on heat transfer in non-Newtonian nanofluid},
  author={Akbari, Omid Ali and Toghraie, Davood and Karimipour, Arash and Marzban, Ali and Ahmadi, Gholam Reza},
  journal={Physica E: Low-Dimensional Systems and Nanostructures},
  volume={86},
  pages={68--75},
  year={2017},
  publisher={Elsevier}
}

@article{venthan2021theoretical,
  title={Theoretical analysis of the heat transfer effect of viscoplastic nanofluids in process intensified chemical systems},
  author={Venthan, S Mullai and Amalraj, I Jayakaran and Kumar, P Senthil},
  journal={Chemical Engineering and Processing-Process Intensification},
  volume={159},
  pages={108227},
  year={2021},
  publisher={Elsevier}
}

@article{ahmed2025numerical,
  title={Numerical investigation on heat and mass transport enhancement in viscoplastic hybrid nanofluid using Bingham-Papanastasiou rheological theory},
  author={Ahmed, M and Nawaz, M},
  journal={International Communications in Heat and Mass Transfer},
  volume={168},
  pages={109448},
  year={2025},
  publisher={Elsevier}
}

@article{hussain2025thermal,
  title={Thermal radiation effects on Bingham--Papanastasiou tetra-hybrid nanofluid dynamics in a porous diseased artery with electro-osmotic flow and permeability},
  author={Hussain, Syed M and Nazar, T and Shabbir, MS and Qureshi, Muhammad Amer and Jamshed, Wasim and Makhdoum, Basim M and Guedri, Kamel and Almaliki, Abdulrazak H and Bayram, Mustafa},
  journal={Journal of Radiation Research and Applied Sciences},
  volume={18},
  number={2},
  pages={101575},
  year={2025},
  publisher={Elsevier}
}

@article{ouyahia2017numerical,
  title={Numerical study of the flow in a square cavity filled with Carbopol-TiO2 nanofluid},
  author={Ouyahia, Seif-Eddine and Benkahla, Youb Khaled and Berabou, Welid and Boudiaf, Ahlem},
  journal={Powder Technology},
  volume={311},
  pages={101--111},
  year={2017},
  publisher={Elsevier}
}

@article{heris2006experimental,
  title={Experimental investigation of oxide nanofluids laminar flow convective heat transfer},
  author={Heris, S Zeinali and Etemad, S Gh and Esfahany, M Nasr},
  journal={International communications in heat and mass transfer},
  volume={33},
  number={4},
  pages={529--535},
  year={2006},
  publisher={Elsevier}
}

@article{srilatha2023heat,
  title={Heat transfer analysis in magnetohydrodynamic nanofluid flow induced by a rotating rough disk with non-Fourier heat flux: aspects of modified Maxwell--Bruggeman and Krieger--Dougherty models},
  author={Srilatha, Pudhari and Madhu, J and Khan, Umair and Kumar, R Naveen and Gowda, RJ Punith and Ahmed, Samia Ben and Kumar, Raman},
  journal={Nanoscale Advances},
  volume={5},
  number={21},
  pages={5941--5951},
  year={2023},
  publisher={Royal Society of Chemistry}
}

@article{alsulami2023three,
  title={Three-dimensional swirling flow of nanofluid with nanoparticle aggregation kinematics using modified Krieger--Dougherty and Maxwell--Bruggeman models: a finite element solution},
  author={Alsulami, MD and Abdulrahman, Amal and Kumar, R Naveen and Punith Gowda, RJ and Prasannakumara, BC},
  journal={Mathematics},
  volume={11},
  number={9},
  pages={2081},
  year={2023},
  publisher={MDPI}
}

@article{sarma2025comparative,
  title={Comparative numerical study of graphene--copper Boger hybrid nanofluid flow with and without aggregation under Robin-type thermal constraints},
  author={Sarma, Neelav and Parasar, Rimjhim},
  journal={Journal of Taibah University for Science},
  volume={19},
  number={1},
  pages={2569762},
  year={2025},
  publisher={Taylor \& Francis}
}

@article{anitha2025impact,
  title={Impact of nanoparticles’ aggregation in a convective flow of non-Newtonian nanofluid in a microchannel: irreversibility analysis},
  author={Anitha, L and J. Gireesha, B},
  journal={International Journal of Ambient Energy},
  volume={46},
  number={1},
  pages={2558153},
  year={2025},
  publisher={Taylor \& Francis}
}

@article{jan2025enhanced,
  title={Enhanced thermal performance of aggregated TiO₂ nanoparticles in non-Newtonian fluid flow over curved surfaces under magnetic and porous medium effects},
  author={Jan, Ahmed and Ahmad, Adeel and Aslam, Farida},
  journal={Journal of Thermal Analysis and Calorimetry},
  volume={150},
  number={18},
  pages={14517--14529},
  year={2025},
  publisher={Springer}
}

@article{article,
author = {Soto, Hilda and Martins-Costa, Maria and Fonseca, Cleiton and Frey, Sergio},
year = {2010},
month = {12},
pages = {450-460},
title = {A numerical investigation of inertia flows of Bingham-Papanastasiou fluids by an extra stress-pressure-velocity galerkin least-squares method},
volume = {32},
journal = {Journal of the Brazilian Society of Mechanical Sciences and Engineering}
}

@article{benkhedda2020convective,
  title={Convective heat transfer performance of hybrid nanofluid in a horizontal pipe considering nanoparticles shapes effect: M. Benkhedda et al.},
  author={Benkhedda, Mohammed and Boufendi, Toufik and Tayebi, Tahar and Chamkha, Ali J},
  journal={Journal of Thermal analysis and Calorimetry},
  volume={140},
  number={1},
  pages={411--425},
  year={2020},
  publisher={Springer}
}

@article{muhammad2025flow,
  title={Flow and heat transfer analysis of hybrid nanofluid due to coaxial cylinders: A comparative study via Keller-box scheme},
  author={Muhammad, Khursheed and Sarfraz, Mahnoor and Aigo, Musa Adam},
  journal={Proceedings of the institution of mechanical engineers, part N: Journal of nanomaterials, nanoengineering and nanosystems},
  pages={23977914251333298},
  year={2025},
  publisher={SAGE Publications Sage UK: London, England}
}

@article{soares1999heat,
  title={Heat transfer to viscoplastic materials flowing laminarly in the entrance region of tubes},
  author={Soares, M{\'a}rcia and Naccache, M{\^o}nica F and Mendes, Paulo R Souza},
  journal={International journal of heat and fluid flow},
  volume={20},
  number={1},
  pages={60--67},
  year={1999},
  publisher={Elsevier}
}

@article{celik2008procedure,
  title={Procedure for estimation and reporting of uncertainty due to discretization in CFD applications},
  journal={Journal of Fluids Engineering},
  author={Celik, Ismail B and Ghia, Urmila and Roache, Patrick J and Freitas, Christopher J and Coleman, Hugh and Raad, Peter E},
  year={2008}
}

@article{khan2026computational,
  title={Computational study of thermal risk mitigation via fractional MHD nanofluid modeling},
  author={Khan, Mumtaz and Anwar, M S and Hussain, Zakir and Irfan, M and Muhammad, Taseer},
  journal={Journal of Hazardous Materials Advances},
  pages={100995},
  year={2026},
  publisher={Elsevier}
}

@article{khan2026distributed,
  title={Distributed-order time-fractional analysis of non-Newtonian Sisko fluid flow under magneto-thermal effects},
  author={Khan, Mumtaz and Anwar, Muhammad Shoaib},
  journal={Propulsion and Power Research},
  year={2026},
  publisher={Elsevier}
}

@article{irfan2025thermal,
  title={Thermal behavior and efficiency enhancement of CuO-Al2O3 hybrid nanofluids using fractional calculus},
  author={Irfan, M and Anwar, M S and Abas, Siti Sabariah and Muhammad, Taseer and Hussain, Zakir and Khan, Mumtaz},
  journal={Journal of Thermal Analysis and Calorimetry},
  volume={150},
  number={3},
  pages={2181--2194},
  year={2025},
  publisher={Springer}
}

@article{muhammad2025significance,
  title={Significance of Brownian diffusion and thermophoresis in energy and mass optimization for Newtonian and Non-Newtonian fluid flow: A numerical study via Keller-Box method},
  author={Muhammad, Khursheed and Sarfraz, Mahnoor and Alrihieli, Haifaa F and Elseesy, Ibrahim E},
  journal={Case Studies in Thermal Engineering},
  volume={72},
  pages={106365},
  year={2025},
  publisher={Elsevier}
}





\end{document}